\documentclass[twocolumn,secnumarabic,amssymb, nobibnotes, nofootinbib,aps, prd, superscriptaddress]{revtex4-1}

\usepackage{dcolumn}
\usepackage{amsmath}
\usepackage{graphicx}
\usepackage{epsf}
\usepackage{psfrag}
\usepackage{pifont}
\usepackage{epsfig}
\usepackage{graphics}
\usepackage{amsfonts}
\usepackage{epstopdf}
\usepackage{slashed}
\usepackage{amsmath,amssymb}
\usepackage{amsthm}
\usepackage{braket}
\usepackage{multirow}
\usepackage{colortbl}

\newcommand{\g}{\gamma}

\newcommand{\s}{\slashed}

\newcommand{\be}{\begin{equation}}
\newcommand{\ee}{\end{equation}}
\newcommand{\bq}{\begin{eqnarray}}
\newcommand{\eq}{\end{eqnarray}}
\newcommand{\ba}{\begin{align}}
\newcommand{\ea}{\end{align}}

\newlength{\ml}
\begin{document}

\title{Supercurrent anomaly and gauge invariance in N=1 supersymmetric Yang-Mills theory}%

\author{Y. R. Batista}
\email{yurirod@fisica.ufmg.br}
\affiliation{UFMG - Universidade Federal de Minas Gerais, ICEx, Dep. de F\'{\i}sica, Av. Ant\^onio Carlos, 6627, Belo Horizonte, MG, Brasil, CEP 31270-901}

\author{Brigitte Hiller}
\email{brigitte@fis.uc.pt}
\affiliation{CFisUC, Department of Physics, University of Coimbra, 3004-516 Coimbra, Portugal}

\author{Adriano Cherchiglia}
\email{adriano.cherchiglia@ufabc.edu.br}
\affiliation{UFABC - Universidade Federal do ABC, CCNH - Centro de Ci\^encias Naturais e Humanas,
Campus Santo Andr\'e, Avenida dos Estados, 5001, Santo Andr\'e - SP - Brasil, CEP 09210-580}

\author{Marcos Sampaio}
\email{marcos.sampaio@ufabc.edu.br}
\affiliation{UFABC - Universidade Federal do ABC, CCNH - Centro de Ci\^encias Naturais e Humanas,
Campus Santo Andr\'e, Avenida dos Estados, 5001, Santo Andr\'e - SP - Brasil, CEP 09210-580}

\begin{abstract}
We analyse Feynman diagram calculational issues related to the quantum breaking of supercurrent conservation in a supersymmetric non-abelian Yang-Mills theory. For the sake of simplicity, we take a zero mass gauge field multiplet interacting with a massless Majorana spin-$1/2$ field in the adjoint representation of $SU(2)$. We shed light on a long-standing controversy regarding the perturbative evaluation of the supercurrent anomaly in connection with gauge and superconformal symmetry in different frameworks. We find that only superconformal symmetry is unambiguously broken using an invariant four dimensional regularization and compare with the triangle AVV anomaly. Subtleties related to momentum routing invariance in the loops of diagrams and Clifford algebra evaluation inside divergent integrals are also discussed in connection with finite and undetermined quantities in Feynman amplitudes.

\end{abstract}


\maketitle

\section{Introduction}

Anomaly-mediated symmetry breaking is an important mechanism in field and string theory \cite{ADLER}. Its range of applications run from phenomenological, such as the
calculation of the decay rate for neutral pions into two photons \cite{ABBJ}, the computation of quantum
numbers in the Skyrme model of hadrons \cite{EBERT} and  mechanisms for baryogenesis in
the Standard Model \cite{BERN}, to theoretical, namely the study of dualities in gauge theory, the computation of anomalous couplings in the effective theory of D-branes, and the analysis of Black Hole entropy \cite{HARVEY, BERTLMANN}.

In the particular case of supersymmetry  breaking, it is neither straightforward nor conclusive that supersymmetry is a symmetry of the full quantum theory in general. 
 However, as discussed in \cite{JJ}, there have been claims about  supersymmetry anomalies which turned out erroneous because of the difficulty to distinguish between a genuine and a spurious anomaly. The latter is  an apparent violation of a supersymmetric Ward identity due to use of a regularization method that violates supersymmetry, for instance. 
 
 The existence of anomalies may be established in a regularization independent way. The Adler-Bardeen anomaly, for instance, can be shown to be determined   by the topological
 term $\mbox{Tr}\, G \tilde{G}$
 algebraically characterized as a non-variation under gauge
 and BRS symmetry. In \cite{KRAUSAN}  was proved that the coefficient of the
 anomaly is determined by convergent  one-loop integrals. Moreover, in \cite{KRAUSSU} it was shown that, with  local  coupling,  supersymmetric  Yang-Mills  theories  have  an
 anomalous  breaking  of  supersymmetry  at  one-loop  order.

Perturbative evaluation of a quantum symmetry breaking is therefore intimately related to regularization issues. A specious anomaly stemming from  non-invariant regularizations appears when finite and regularization dependent terms are erroneously incorporated into an amplitude. A symmetry preserving regularization is of considerable computational utility. Evidently if a model is known beforehand  to be anomaly free, the question of whether there exists or not an invariant scheme is irrelevant, should the imposition of Ward identities order by order in perturbation theory not to be considered a nuisance. In this case either, one employs an invariant scheme and performs renormalization using invariant counterterms or uses non-invariant counterterms to compensate the symmetry breaking.  For the latter strategy to work, a precise knowledge of the symmetry content of the model must be known which often requires non-perturbative \cite{SCHWARZ} information. In fact, the absence of anomalies may be proven without recourse to any regularization by using algebraic properties of the Ward identities (algebraic renormalization) at least for some particular cases \cite{JJ,KRAUS,BECCHI,WHITE,PIGUET}.

Although dimension regularization  \cite{DR} is tailor-made for gauge theories,  it is less suited to dimensional sensitive quantum field theoretical models such as supersymmetric, topological and chiral gauge theories. Naive dimensional reduction (DRed) can be shown mathematically inconsistent \cite{SIEGEL}. For instance, Lorentz algebra contractions can lead to equations such as $0=n(n-1)(n-2)(n-3)(n-4)$, valid only for integer spacetime dimension $n$ and thus it is incompatible with analytical continuation proposed by dimensional methods.
In \cite{STOCKINGERQAP}, a consistent version of DRed was developed which forbids the use of Fierz identities, implying that supersymmetry will also not be respected in general. Nevertheless, in the same reference it is shown how to identify the breaking of supersymmetry by means of the quantum action principle, allowing DRed to
be made operational for particular models
to a specific loop order \cite{STOCKINGERNPB}. 
In the same vein,
there were severe difficulties to renormalize supersymmetric theories in a regularization independent way: in the Wess-Zumino gauge useful in practical calculations, the usual way of treating global symmetries by Ward identities was shown to fail \cite{MAISON,STOCKINGERQAP}. 

Anomaly mediated supersymmetry breaking is an important  mechanism in addition to Planck scale mediated and gauge mediated scenarios \cite{JUNG}. The former is related to superconformal anomaly and started  back in early 1970’s with developments in rigid supersymmetry models \cite{FZSS}. The study of supersymmetry breaking involves the computation of Ward identities connecting Greens functions of the supercurrent to other matrix elements. For those Ward identities to hold, no anomalies of the supercurrent should exist. Although a manisfest gauge and supersymmetry invariant regularization is still to be constructed some  regularization frameworks that operate in the physical dimension do exist \cite{AGUILA}-\cite{GERMAN}. In particular, Implicit Regularization (IReg), developed by Batisttel and collaborators, \cite{ORIMAR}-\cite{JOILSON} systematically identifies, to arbitrary loop order, regularization dependent terms as surface terms (resulting from differences between loop integrals free of external momenta with the same superficial degree of divergence)  without recourse to an explicit regulator. For a comparison with other similar emergent schemes, see \cite{ZURICH}. IReg has been shown to be adequate to connect momentum routing invariance in a  diagram, gauge invariance and surface terms in the corresponding Feynman amplitude and therefore it will be used as a tool in this contribution.  

We revisit an old controversy regarding the diagrammatic evaluation of the supercurrent anomaly that started with de-Witt and Freedman \cite{DWITT}. For concreteness we study $N=1$ super Yang-Mills $SU(2)$ theory in four spacetime dimensions. In this model, the supercurrent, the axial current and the stress-energy tensor belongs to the same multiplet, that is  they transform among themselves under constant supersymmetry transformations. As  the axial-vector current has an anomaly, one is compelled to conclude that the supercurrent conservation could be anomalous as well \cite{ABBOTT1}.\footnote{For a unified discussion about chiral and conformal anomalies see \cite{CORIANO}. Also, see \cite{MehtaN} for a study about path integral derivation of anomalies.} 

The quantum breaking of such symmetry constraints translates into the violation of one amid three Ward identities which hold at classical level. As we shall discuss, different calculational frameworks placed the anomaly in one of the Ward identities. An important result by Abbott, Grisaru and Schnitzer \cite{ABBOTT1,ABBOTT2} shows, however, that one cannot derive the quantum breaking of the supercurrent conservation from the axial-vector current anomaly. Moreover in \cite{MACKEON}, within a four dimensional approach called Pre-regularization, it was shown that the supercurrent anomaly is connected to the inability to reconcile ambiguities (in the form of {{\it specific}} momentum routings in the Feynman diagrams) in a way to preserve simultaneously gauge and supersymmetry. 
 
 The purpose of this work is to shed light on the  apparent clash in the perturbative calculation of the quantum symmetry breaking  of the supercurrent conservation in N=1 super Yang Mills theory in connection with  gauge invariance and the Rarita-Schwinger constraint (the latter also known as superconformal, spin-$3/2$ or supercurrent trace constraint) \cite{ABBOTT1}. Subtleties related to Dirac algebra and symmetric integration within divergent amplitudes, parametrization of arbitrary (finite) regularization dependent terms and momentum routing invariance in a framework which operates in the physical dimension such as IReg will be clarified. Moreover a comparison with results of the supercurrent anomaly performed in other regularization frameworks and a comparison with the  Adler-Bardeen-Bell-Jackiw (ABBJ) triangle anomaly  will be presented.

This contribution is organized as follows. In section II we present some technical tools and apply them to ABBJ anomaly. The Feynman rules of SU(2) super Yang-Mills Lagrangian with massless Majorana spinors in the presence of an external current ${\cal{S_\mu}}$ as well as the one loop correction to the process ${\cal{S_\mu}} \rightarrow \psi + A_\mu $ compose most of section III. It also contains the corresponding Ward identities respected at classical level and subject to quantum breaking. The results of the Ward identities within different schemes with  focus in IReg follow in sections IV and V. Sections VI and VII contain a general discussion and conclusions regarding subtleties appearing in perturbative evaluation of anomalies. All technical details are left to appendices.

\section{IReg and the ABBJ triangle Anomaly}

IReg \cite{ORIMAR}-\cite{JOILSON} is a regularization framework applicable to arbitrary loop order proposed as an alternative to dimensional schemes. It operates in the physical
dimension of the underlying quantum field theory avoiding
some  drawbacks of dimensional methods, for instance the mismatch between fermionic and
bosonic degrees of freedom that leads to  supersymmetry breaking and ambiguities in the $\gamma_5$ matrix and Levi-Civita tensor algebra due to dimensional continuation on the spacetime dimension. Moreover, spurious evaluation of regularization dependent parameters \cite{JACKIWFU}, which is usual in some non-dimensional frameworks,
are avoided.
IReg operates in momentum space using as main strategy the isolation of basic divergent loop integrals (BDIs) of a given superficial degree of divergence  that characterizes the UV divergent behaviour of  Feynman amplitudes. The latter is freed from external momentum dependence by judiciously applying an algebraic identity at the integrand level:
\begin{align}
\frac{1}{(k-p)^2-m^2}
=\frac{1}{(k^2-m^2)}
+\frac{(-1)(p^2-2\,p \cdot k)}{(k^2-m^2)\left[(k-p)^2-m^2\right]}\,.
\label{id}
\end{align}
It resembles in some aspects others four dimensional programs such as differential renormalization \cite{AGUILA,Sampaio2002} and the FDR scheme \cite{ZURICH,PITTAU} in which the intrinsic divergent pieces are called ``vacua". Infrared divergences can be regulated either by a fictitious mass at propagator level or by infrared basic integrals in coordinate space \cite{IRI}. Tensorial basic divergent integrals in turn may be expressed as scalar ones plus surface terms (ST). STs encode most of the regularization dependent pieces of explicit regularizations. These features are especially useful for dimensional specific quantum field theories. Moreover the IReg scheme can be generalized to arbitrary loop order complying with the BPHZ renormalization program \cite{BPHZ}. After subtraction of subdivergences following the Bogoliubov's recursion formula (devised for subtracting nested and overlapping divergences) it is still possible to define new  BDIs and surface terms characterizing the divergent behaviour at arbitrary loop order. Here, for the sake of brevity, we shall describe only the one loop structure of IReg.

To establish our notation, we write one loop logarithmically basic divergent integrals as \footnote{$\int_k \equiv \int^\Lambda d^4k/(2 \pi)^4$, $\Lambda$ being a $4$-dimensional implicit regulator (say, a cutoff) just to justify algebraic operations within the integrands}:

\begin{equation}
I^{\mu_1 \cdots \mu_{2n}}_{0}(m^2)\equiv \int_k \frac{k^{\mu_1}\cdots k^{\mu_{2n}}}{(k^2-m^2)^{2+n}},
\end{equation}
with similar definitions for linearly and quadratically divergent objects. One loop STs  are defined by

\begin{eqnarray}
\Upsilon^{\mu \nu}_{2w} &=&  g^{\mu \nu}I_{2w}(m^2)-2(2-w)I^{\mu \nu}_{2w}(m^2)  \equiv \upsilon_{2w}g^{\mu \nu},
\label{dif1} \\
\Xi^{\mu \nu \alpha \beta}_{2w} &=&  g^{\{ \mu \nu} g^{ \alpha \beta \}}I_{2w}(m^2)
 -4(3-w)(2-w)I^{\mu \nu \alpha \beta }_{2w}(m^2) \nonumber \\ &\equiv&
\xi_{2w}(g^{\mu \nu} g^{\alpha \beta}+g^{\mu \alpha} g^{\nu \beta}+g^{\mu \beta} g^{\nu \alpha}),
\label{dif2}
\end{eqnarray}
\noindent 
etc., $2w$ being the degree of divergence of the integrals (hereafter we identify the subscripts $0, 1, 2$ with
$log$, $lin$, $quad$). The curly brackets above stand for symmetrization in the Lorentz indices. It is straightforward to show that STs are integrals of total derivatives, 
\begin{eqnarray}
\upsilon_{2w}g^{\mu \nu}= \int_k\frac{\partial}{\partial k_{\nu}}\frac{k^{\mu}}{(k^2-m^2)^{2-w}},
\label{ts1}
\end{eqnarray}
\begin{eqnarray}
&&(\xi_{2w}-v_{2w})(g^{\mu \nu} g^{\alpha \beta}+g^{\mu \alpha} g^{\nu \beta}+g^{\mu \beta} g^{\nu \alpha})=\nonumber \\ && \int_k\frac{\partial}{\partial 
k_{\nu}}\frac{2(2-w)k^{\mu} k^{\alpha} k^{\beta}}{(k^2-m^2)^{3-w}}.
\label{ts2}
\end{eqnarray}
The differences between integrals with the same degree of divergence (\ref{dif1}) and (\ref{dif2}) are regularization dependent and should be fixed  by symmetry constraints or phenomenology \cite{JACKIWFU}.  As shown for instance in \cite{Adriano2012}, such STs are intimately connected with  momentum routing invariance (MRI) in the loops of a Feynman diagram. By  consistently setting STs to zero order by order in perturbation theory enforces both MRI and  gauge invariance \cite{Viglioni2016,Adriano2012}, allowing us to
conjecture that STs are at the root of some symmetry breakings in Feynman diagram calculations. For instance in  \cite{Adriano2012,Ottoni2006,FARGNOLISUSY,CHERCHIGLIAEPJC2016} it was shown that constrained IReg (that is, systematically setting STs to vanish) is also a necessary condition for 
supersymmetry invariance. Similar results using different theories were obtained for non-abelian gauge theories  \cite{Cherchiglia2014,Pontes,FARGNOLISUSY}. More recently, IReg was shown to be useful in dealing with $\gamma_5$ algebra issues in Feynman amplitudes \cite{JOILSON,ADRIANOPEREZ}. 
 
Finally, a mass dimensional parameter in BDIs can be extracted to define a minimum and mass independent subtraction scheme via  a regularization independent identity which, at one loop order, reads
\begin{align}
I_{{log}}(m^2)
=I_{{log}}(\lambda^2)-b\,\ln \left(\frac{m^2}{\lambda^2}\right)\,,
\label{SR}
\end{align}
where 
\be
b \equiv \frac{i}{(4 \pi)^2}
\label{b}
\ee
and $\lambda$ plays the role of a renormalization group scale. Derivatives of BDIs with respect to $\lambda$ yield a constant or another BDI. They are absorbed into renormalization constants without explicit evaluation allowing the computation of the usual renormalization group functions.


\subsection{ABBJ anomaly within IREG}

As an illustration which will serve to interpret the results  of the supercurrent anomaly, let us  present a classical example, namely the Adler-Bardeen-Bell-Jackiw (ABBJ) triangle anomaly. Calculational details have been presented elsewhere \cite{Viglioni2016,Adriano2012}. Since its discovery \cite{ABBJ}, the ABBJ anomaly has been calculated within several approaches \cite{MACKEON},\cite{YU}-\cite{DALLABONA}. An overview on the various regularization
schemes applied in the diagrammatic computation of this anomaly can be found in \cite{BERTLMANN}. There 
are also derivations obtained by the path integral measure transformation 
\cite{FUJIKAWA} and differential geometry \cite{ZEE,BARDEENOV}. The usual view on the diagrammatic derivation of chiral anomaly  is that it  
occurs due to the momentum routing breaking in the internal lines of Feynman diagrams. Accordingly, the momenta of the internal lines must assume specific values so that 
a certain Ward identity is preserved \cite{CURRENT}. This feature was formalized in \cite{MACKEON} within a scheme called Preregularization.

It is not uncommon that an anomaly can  apparently be removed and reappear
in another guise. This is indeed the case with the ABBJ anomaly, which is a property of the fermion triangle with two vector and one axial–vector vertices. The anomaly may affect either the axial  or the vector current, depending on how the theory is regularized. However it is not up to the regularization scheme which identity is to be preserved. Ideally the calculational framework should
democratically display the anomaly which, if not spurious, contains important physical implications. Besides the triangle anomaly, other examples of interchangeable anomalies  are gravitational anomalies for fermions in $2$-dimensional spacetime \cite{Souza2005}. The supercurrent anomaly was thought to have the same property but as we shall discuss in this contribution, at least within a diagrammatic evaluation, this is not the case. 

Standard dimensional regularization \cite{DR} is the most  appropriate invariant method for vector gauge invariance. However, as we have mentioned above, some inconsistencies may appear regarding the manipulation of dimension specific objects such as $\gamma_5$-matrix and the Levi-Civita tensor. To circumvent this problem, some rules had to be added to the method, postulating how the dimensional continuation of such objects should be performed \cite{JEGERLEHNER}. Dimensional methods also break supersymmetry and some ad-hoc rules must be incorporated as well. To this end, some supersymmetric Ward identities of the underlying model have to be validated  to a certain loop order via say, the quantum action principle \cite{STOCKINGERQAP}. Alternatively, a strategy that  imposes  Ward identities to restore broken symmetries order by order in perturbation theory could be employed. However, besides being not practical from the calculational viewpoint, it can also be misleading when there exists a genuine anomaly.

In the particular case of the AVV anomaly, the triangle graph contains an axial vertex and care must be exercised with divergent amplitudes involving the dimension specific object $\gamma_5$-matrix and its Clifford Algebra. That is because identities regarding the $\gamma_5$ algebra are not always satisfied under divergent integrals, even in the physical dimension of the model \cite{Viglioni2016},\cite{ADRIANOPEREZ}. A gauge-invariant prescription for the 
$\gamma_5$ algebra was proposed in \cite{TSAI} called Rightmost Ordering in which all $\gamma_5$ should be moved to the rightmost position of the amplitude before its space-time dimensionality is altered.  Another proposal in four dimensions  was discussed in\cite{CYNOLTER}. Moreover,  in \cite{ADRIANOPEREZ}, it can be found a thorough discussion on calculations involving Clifford algebra within Feynman amplitudes evaluated in different schemes .  

Although  the evaluation of the ABBJ anomaly has  been extensively discussed in the literature, we briefly recall some aspects of its calculation within IReg with the purpose of comparing with the supercurrent anomaly. The key features are firstly the parametrization of regularization dependent (and undetermined) quantities as surface terms and secondly MRI in the loops of a Feynman amplitude while working in the physical dimension where the Clifford algebra is defined. MRI is known to be fulfilled in  cases where gauge symmetry is not broken at all orders in perturbation theory upon the use of Dimensional Regularization \cite{DR}.

It is legitimate to expect that even in the presence of an anomaly, vector current gauge invariance continues to evidence MRI which is built up from energy-momentum conservation at a diagram vertices . Indeed that is what we verify by applying a minimal prescription  based on the symmetrization of the trace over the $\gamma$ matrices involving $\gamma_5$. This prescription does not make use of the property $\{\gamma_5,\gamma_\mu \}=0$, since  the vanishing of such anti-commutator inside divergent integrals is  the origin of ambiguities \cite{Viglioni2016} even when applied in the physical dimension \cite{ADRIANOPEREZ,SIGNER5}. Thus, for the traces involving four and six Dirac matrices and one $\gamma_5$ we use:
\be
\mbox{Tr}[\gamma^\mu \gamma^\beta \gamma^\nu \gamma^\rho \gamma^5] = 4 \,\, i\,\, \epsilon^{\mu \beta \nu \rho} \,\,\,\,\, \mbox{and} 
\label{traco1}
\ee
\bq
&&\frac{-i}{4}Tr[\gamma ^{\mu }\gamma ^{\nu }\gamma ^{\alpha }\gamma ^{\beta }\gamma ^{\gamma }\gamma ^{\delta }\gamma ^5]=  \nonumber \\ &&
-g^{\alpha \beta } \epsilon ^{\gamma \delta \mu \nu }+g^{\alpha \gamma } \epsilon ^{\beta \delta \mu \nu }-g^{\alpha \delta } \epsilon ^{\beta 
\gamma \mu \nu }-g^{\alpha \mu } \epsilon ^{\beta \gamma \delta \nu } \nonumber \\ &&+ g^{\alpha \nu } \epsilon ^{\beta \gamma \delta \mu } -g^{\beta 
\gamma } \epsilon ^{\alpha \delta \mu \nu }+g^{\beta \delta } \epsilon ^{\alpha \gamma \mu \nu }+g^{\beta \mu } \epsilon ^{
\alpha \gamma \delta \nu } \nonumber \\ &&-g^{\beta \nu } \epsilon ^{\alpha \gamma \delta \mu }-g^{\gamma \delta } \epsilon ^{\alpha \beta \mu \nu }
-g^{\gamma \mu } \epsilon ^{\alpha \beta \delta \nu }+g^{\gamma \nu } \epsilon ^{\alpha \beta \delta \mu } \nonumber \\ && +g^{\delta \mu } 
\epsilon ^{\alpha \beta \gamma \nu }-g^{\delta \nu } \epsilon ^{\alpha \beta \gamma \mu }-g^{\mu \nu } \epsilon ^{\alpha \beta \gamma \delta }, \label{traco2}
\eq
which can be obtained replacing $\gamma_5$ by its definition in four spacetime dimensions, 
$\gamma_5=\frac{i}{4!}\epsilon^{\mu\nu\alpha\beta}\gamma_{\mu}\gamma_{\nu}\gamma_{\alpha}\gamma_{\beta}$. A similar approach as encoded in equation (\ref{traco2}) was used in \cite{CYNOLTER,ADRIANOPEREZ}. Notice that if we had used the following identity to reduce the number of Dirac matrices,
\be
\gamma ^{\mu }\gamma ^{\beta }\gamma ^{\nu }= g^{\mu \beta} \gamma^{\nu}+g^{\nu \beta} \gamma^{\mu}-g^{\mu \nu} \gamma^{\beta}-i 
\epsilon^{\mu \beta \nu \rho}\gamma_{\rho}\gamma_5,
\label{gamma}
\ee
then both equation (\ref{traco1})
and the anti-commuting property $\gamma_5\gamma^{\rho}\gamma_5=-\gamma^{\rho}$ would lead to
\bq
&& Tr[\gamma ^{\mu }\gamma ^{\beta }\gamma ^{\nu }\gamma ^{\xi }\gamma ^{\alpha}\gamma ^{\lambda }\gamma ^5]= 4i\big (g^{\beta \mu}\epsilon^{\nu 
\xi \alpha \lambda}+g^{\beta \nu}\epsilon^{\mu \xi \alpha \lambda} \nonumber \\ &&-g^{\mu \nu}\epsilon^{\beta \xi \alpha \lambda}
-g^{\lambda \alpha}\epsilon^{\mu \beta \nu \xi}+g^{\xi \lambda}\epsilon^{\mu \beta \nu \alpha}-g^{\xi \alpha}\epsilon^{\mu \beta \nu 
\lambda}\big).
\label{trace}
\eq
However, it is completely arbitrary which three $\gamma$ matrices should be taken in order to apply equation (\ref{gamma}). A different choice would give  equation (\ref{trace}) with Lorentz indexes permuted. Such arbitrariness turns out to be relevant  in connection with symmetry breakings. Thus, the anti-commutation property $\{ \gamma_\mu ,\gamma_5 \} = 0$ should be avoided inside a divergent integral. Moreover, it has been shown that  this operation can  assign an a priori non-vanishing value to an arbitrary surface term \cite{Viglioni2016}.

\begin{figure*}[!htb]
\includegraphics[trim=0mm 100mm 0mm 100mm,scale=0.5]{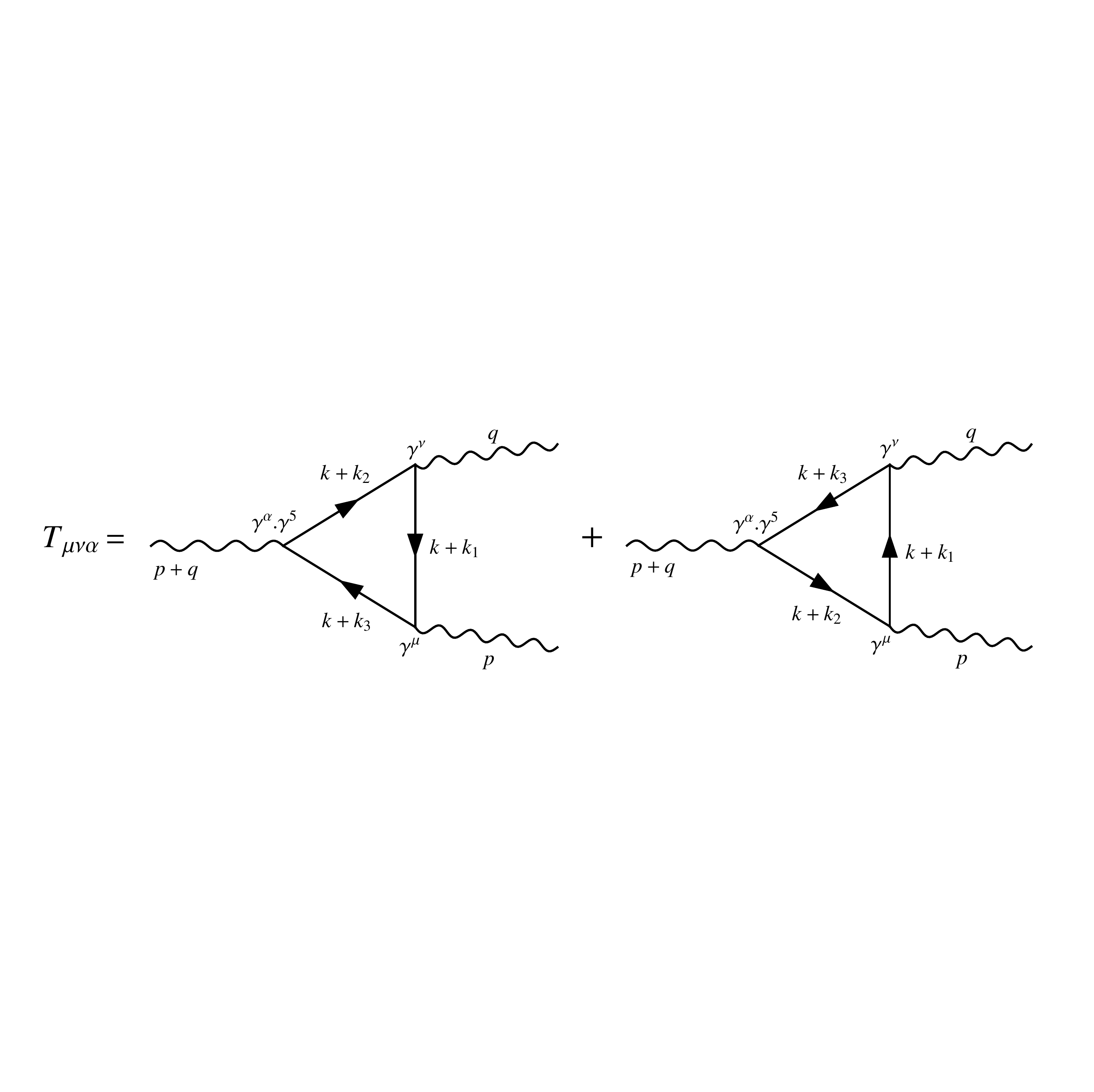}
\caption{Triangle diagrams which contribute to the ABBJ anomaly. We label the internal lines with arbitrary momentum routings.}
\label{triangle}
\end{figure*}
The amplitude of the Feynman diagrams in figure \ref{triangle} is given by
\bq
T_{\mu\nu\alpha}&=&-i\int_k {\mbox{Tr}}\Big[\gamma_{\mu}\frac{i}{\slashed{k}+\slashed{k}_1-m}\gamma_{\nu}\frac{i}{\slashed{k}+\slashed{k}_2-m} \nonumber \\
&\times& \gamma_{\alpha}\gamma_5 \frac{i}{\slashed{k}+\slashed{k}_3-m}\Big]+(\mu \leftrightarrow \nu, p\leftrightarrow q).
\label{AVV}
\eq
where the arbitrary routings $k_i$´s obey the following relations due to energy-momentum conservation at each vertex
\begin{align}
k_2-k_3=& p+q,\nonumber\\
k_1-k_3=& p,\nonumber\\
k_2-k_1=& q.
\label{routing}
\end{align}
Equations (\ref{routing}) allow us to parametrize the routing  $k_i$ as
\begin{align}
k_1=& \alpha p+(\beta-1) q,\nonumber\\
k_2=& \alpha p+\beta q,\nonumber\\
k_3=& (\alpha-1) p+(\beta-1) q,
\label{routing2}
\end{align}
where $\alpha$ and $\beta$ are arbitrary real numbers which express the freedom to choose the routing of internal lines. Similar equations  for the other diagram are obtained by interchanging $ p\leftrightarrow q$. After taking the trace with the help of equation (\ref{traco2}), we apply the IReg scheme to obtain
\be
T_{\mu\nu\alpha}= 4i\upsilon_0 (\alpha-\beta-1)\epsilon_{\mu \nu \alpha \beta}(q-p)^{\beta}+ \widetilde{T}_{\mu\nu\alpha},
\label{resultado}
\ee
where $\upsilon_0$ is a surface term (generally explicitly evaluated in other regularization schemes) as defined in  (\ref{ts1}) and $\widetilde{T}_{\mu\nu\alpha}$ is the finite part of the amplitude sketched in appendix D.  The vector and axial Ward identities for the massless theory read:
\begin{align}
&p_{\mu}T^{\mu \nu \alpha}=-4i\upsilon_0 (\alpha-\beta-1)\epsilon^{\alpha \nu \beta \lambda}p_{\beta}q_{\lambda},\nonumber\\
&q_{\nu}T^{\mu \nu \alpha}=4i\upsilon_0 (\alpha-\beta-1)\epsilon^{\alpha \mu \beta \lambda}p_{\beta}q_{\lambda},\nonumber\\
&(p+q)_{\alpha}T^{\mu \nu \alpha}=8i\upsilon_0 (\alpha-\beta-1)\epsilon^{\mu \nu \beta \lambda}p_{\beta}q_{\lambda} -\frac
{\epsilon^{\mu \nu \beta \lambda}}{2\pi^2}p_{\beta}q_{\lambda}.
\label{WI}
\end{align}
The number $\upsilon_0 (\alpha-\beta-1)$ is arbitrary since $\upsilon_0$ is a (finite) difference between
two logarithmic divergences and $\alpha$ and $\beta$ are  real numbers that we are free to choose as long as equations (\ref{routing}) representing  energy-momentum conservation hold. We can parametrize this arbitrariness in a single parameter $a$ by redefining 
$4i\upsilon_0(\alpha-\beta-1)\equiv\frac{1}{4\pi^2}(1+a)$. Then the Ward identities become
\begin{align}
&p_{\mu}T^{\mu \nu \alpha}=-\frac{1}{4\pi^2}(1+a)\epsilon^{\alpha \nu \beta \lambda}p_{\beta}q_{\lambda},\nonumber\\
&q_{\nu}T^{\mu \nu \alpha}=\frac{1}{4\pi^2}(1+a)\epsilon^{\alpha \mu \beta \lambda}p_{\beta}q_{\lambda},\nonumber\\
&(p+q)_{\alpha}T^{\mu \nu \alpha}=\frac{1}{2\pi^2}a\epsilon^{\mu \nu \beta \lambda}p_{\beta}q_{\lambda}.
\label{WI22}
\end{align}
\noindent
Notice that the anomaly is democratically displayed in the Ward identities (\ref{WI22}). For vector gauge invariance to be  preserved, one chooses $a=-1$ and thus the axial identity is violated by a quantity equal to $-\frac{1}{2\pi^2}\epsilon^{\mu \nu \beta \lambda}p_{\beta}q_{\lambda}$. On the other hand, chiral symmetry is maintained at the quantum level for $a=0$, and the vector identities are violated. The choice
$a=-1$ sets  STs to zero \cite{Adriano2012,Viglioni2016}. In these references it was proved that setting STs to zero assures momentum routing invariance and consequently gauge invariance in abelian gauge theories to arbitrary loop order.  In other words MRI in Feynman diagrams is a necessary and sufficient condition for gauge invariance in abelian gauge theories. Although no general proof has been constructed, the same appears to hold for nonabelian gauge invariance in Feynman diagram calculations \cite{Pontes,FARGNOLISUSY}.  In our result it was crucial to take the symmetrized version of traces involving $\gamma_5$ matrix as displayed in (\ref{traco2}) rather than (\ref{trace}) \footnote{ In \cite{BaetaScarpelli20012} the vector current Ward identities were satisfied when the surface term
assumed a nonvanishing value that cancelled a finite term in order to preserve gauge symmetry. This was a byproduct of identity (\ref{trace}) which was used in that calculation. Therefore, it was thought
that the anomaly was due to the breaking of the momentum routing invariance.}.

In summary, we have seen that within a four dimensional regularization scheme such as IReg and taking into account some subtleties related to Clifford algebra inside divergent integrals (symmetrization of the trace), arbitrary momentum routing amounts to gauge invariance. Such a routing arbitrariness in a Feynman graph is always accompanied by a surface term, which is set to zero on gauge  invariance grounds. Indeed even in the case when  the axial Ward identity is  chosen to be verified in the ABBJ anomaly, the momentum routing may be absorbed in the choice of the arbitrary ST. Thus it seems plausible to 
disregard particular momentum routings in the discussion of Ward identities, even in anomalous cases, in favour of intrinsic arbitrary parameters hidden in perturbation theory in the form of STs. The latter should be left as an adjustable parameter if not fixed on symmetry principles of the underlying model \cite{Viglioni2016,JACKIWFU}.

\section{Supercurrent Anomaly in SUSY Non-Abelian Gauge Theory}

The $N=1$ supersymmetric and gauge invariant Yang-Mills SU(2) Lagrangian in the Wess-Zumino gauge reads \cite{FZSS}  
\begin{eqnarray}
 \mathcal{L} &=&  -\frac{1}{4}F^{a}_{\mu \nu}F^{{a} \mu\nu}+\frac{i}{2}\bar{\psi}^{a}\gamma^\mu \left(D_\mu \psi \right)^{a}+\frac{1}{2}C^{a}C^{a}
\nonumber \\
&+& \eta_a^*\partial^\mu D_\mu^{ab} \eta_b + \frac{1}{2 \xi} [\partial_\mu A^\mu_a]^2\;, \nonumber \\
&\equiv& \mathcal{L}_{inv} + \mathcal{L}_{ghost} + \mathcal{L}_{gauge}
\; , \label{lsym}
\end{eqnarray}
where
\begin{equation}
 F^{a}_{\mu \nu}=\partial_\mu A_\nu^{a} - \partial_\nu A_\mu^{a} + g \epsilon_{{abc}}A_\mu^{b} A_\nu^{c} \; ,
\end{equation}
${a}={1, 2, 3}$ are colour indices, $\psi^{a}$ is a massless Majorana spinor (transforming in the adjoint representation), supersymmetric partner of the gauge fields $A_\nu^{c}$ and $C^{a}$  is an auxiliary field which  we will drop out later as it does not couple to any field in the computation of the supercurrent. The covariant derivative is defined as 
\begin{equation}
 \left(D_\mu \psi \right)^{a}=\partial_\mu \psi^{a} + g \epsilon_{abc} A_\mu^{b} \psi^{c} \;.
\end{equation}

The action correspondent to ${\cal{L}}_{inv}$ is invariant under supersymmetry transformations,
\begin{subequations}\label{ttrfs}
  \begin{gather}
    \delta A_\nu^{a} = i \overline{\epsilon} \gamma_\mu \psi^{a} \; , \label{trfs1}\\
    \delta \psi^{{a}}=\sigma^{\mu \nu} \epsilon F^{{a}}_{\mu \nu} + \epsilon C^{a} \; ,\label{trfs2}\\
    \delta C^{a} = \overline{\epsilon}\gamma^\mu D_\mu \psi^{a} \; , \label{trfs3}
  \end{gather}\label{eq:tsusy}
\end{subequations}

\noindent where $\sigma^{\mu \nu}= \frac{1}{4}\left[\gamma^\mu,\gamma^\nu\right]$
and $\epsilon$ is a constant spinor. The Noether current associated to the invariance of (\ref{lsym}) under (\ref{ttrfs}) is
\begin{subequations}
\begin{gather}
 \overline{\mathcal{S}}_\mu \epsilon= -i \left(\overline{\psi}_a \gamma_\mu \sigma^{\alpha \beta} \epsilon\right)F^{a}_{\alpha \beta}  
\end{gather}
  \textnormal{or equivalently, by the Majorana condition for $\epsilon$ and $\psi (x)$}
\begin{gather}
 \overline{\epsilon}\mathcal{S}_\mu= -i \left(\overline{\epsilon}\sigma^{\alpha \beta}\gamma_\mu \psi_{a}\right)F^{a}_{\alpha \beta} \quad \; . 
\end{gather}\label{spc}
\end{subequations}

Likewise, defining global supersymmetric transformations as in (\ref{eq:tsusy}) but substituting $\gamma_\mu$ with $\gamma_\mu \gamma_5$ and $\sigma _{\mu \nu}$ with  $\sigma _{\mu \nu} \gamma_5$ leads to a conserved  Noether current, which is just (\ref{spc}) with
$\gamma_\mu \rightarrow \gamma_\mu \gamma_5$. The result is not affected by  which definition of the transformation one uses as a $\gamma_5$ factor can be absorbed into the transformation properties of the fields \cite{MAJUMDAR} \footnote{Indeed should one employ the supercurrent definition with the $\gamma_5$-matrix  into the Feynman rules,  we see that applying, for instance, the rightmost positioning (gauge invariant) prescription for the $\gamma_5$ matrix \cite{TSAI} leads to the same results for the quantum corrections modulo a $\gamma_5$ factor.}.

In order to study the on mass-shell anomalies of the supercurrent we consider the process
\begin{equation}
{\cal{S}}_\mu \rightarrow \psi + {A}_\mu
\label{process}
\end{equation}
for on-shell bosons and fermions (thereupon, no need to consider mass and wave function renormalization at the one loop level). The Feynman rules for the bubble and triangle graphs that contribute to the process (\ref{process})  required to evaluate quantum breakings of the supercurrent Ward identities
are displayed in figure \ref{fig:rf} in the Feynman gauge.
In \cite{DWITT} it was shown that matrix
elements of this current between physical
states are gauge invariant and conserved:


\begin{figure*}
  \begin{center}
   \includegraphics[height=7.0cm]{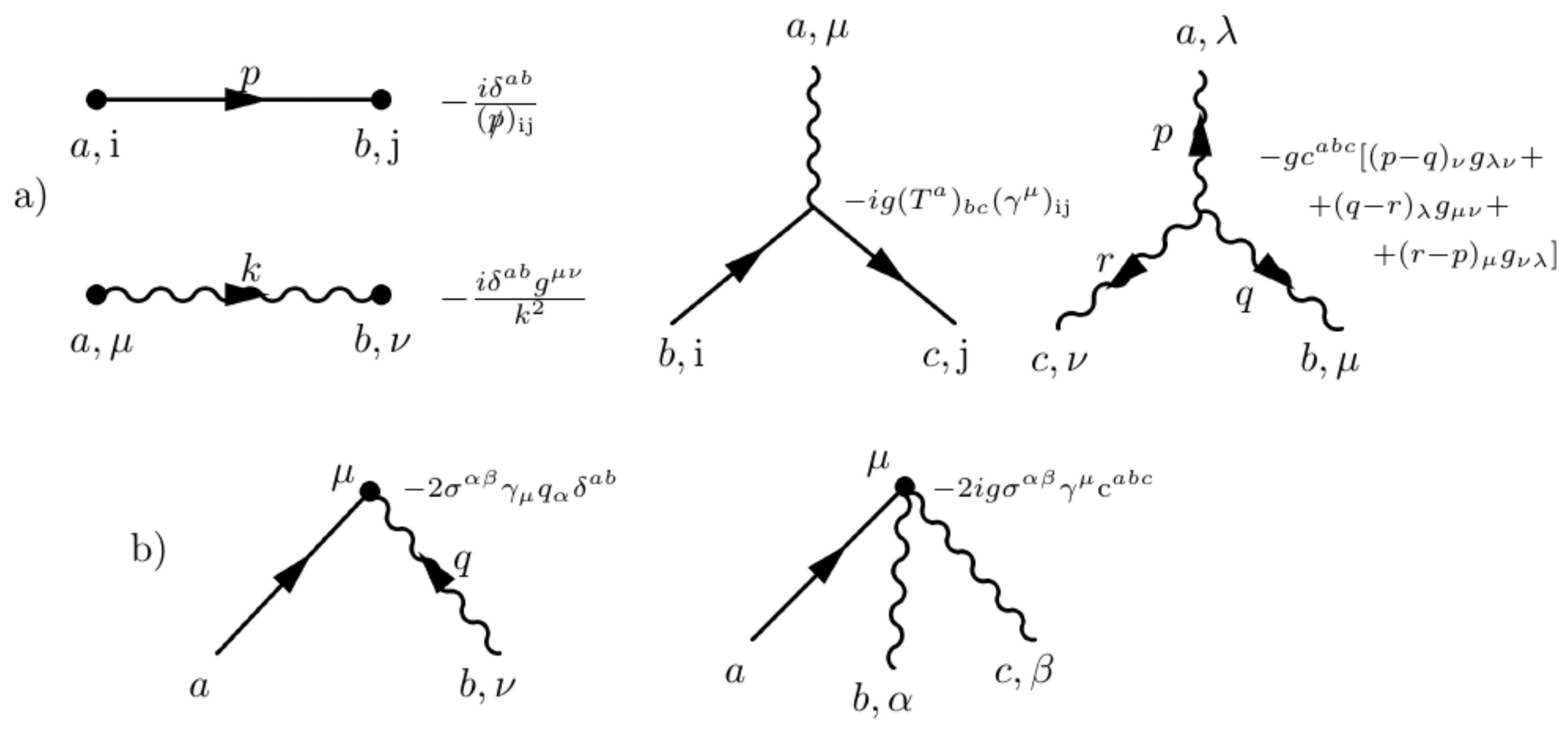}
  \end{center}
\caption{Feynman rules: a) Gauge boson and fermion propagators and interaction vertices b) Supercurrent  $S_{\mu\nu}^{ab}$ vertices} 
\label{fig:rf}
\end{figure*}


\begin{equation}
 \partial_\mu \langle \textnormal{phys}|\mathcal{S}^\mu |\textnormal{phys}\rangle=0 \;.
\end{equation}
The external physical state is a fermion with momentum $p$ and color index $a$, and a gauge field  with momentum $q$ and color index $b$, namely $|u,p,a\rangle | \varepsilon,q,b \rangle$. The transition amplitude to vacuum state reads 
\begin{equation}
 \langle \textnormal{phys}|\mathcal{S}^\mu |\textnormal{phys}\rangle=\langle 0|\mathcal{\bar{S}}_\mu |u,p,a\rangle | \varepsilon,q,b\rangle =  S^{ab}_{\mu \nu} \,\varepsilon^\nu_b(q)\, u_a(p)\;.
\end{equation}
For \textit{on-shell} bosons and fermions all factors of $\s{p}$ in $S_{\mu \nu}^{ab}$ adjacent to $u_a(p)$ vanish as well as factors of $p^2$, $q^2$ or $q_\nu$ in $S_{\mu \nu}^{ab}$:
\bq
    \varepsilon^\nu(q) \,q_\nu = 0\;,\quad \s{p}\,u(p) = 0\;, \quad p^2=q^2=0,
    \label{OSC}
    \eq
in which $\s{p}$ should be moved to the rightmost position to apply (\ref{OSC}).

The Ward identities which express supersymmetry, gauge and superconformal (spin-$3/2$ constraint) invariance of the supercurrent read
\bq
 (p^\mu - q^\mu)S^{ab}_{\mu \nu}&=&0 \; ,\\
  q^\nu S^{ab}_{\mu \nu}&=&0 \; ,\\
  \gamma^\mu S^{ab}_{\mu \nu}&=&0\; ,
  \label{eq:iw}
\eq
where $q_\nu$  is the momentum of the external gauge field. They are readily satisfied at tree level 
\be
(S_{\mu \nu}^{{ab}})^{{{tree}}} = -2 i \sigma_{\beta \nu} \gamma_{\mu} q^{\beta} \delta^{{ab}}\, ,
\ee
but an anomaly occurs at quantum level. Many authors have calculated this anomaly and there has been some controversy about whether the anomaly is in the divergence or in the trace of the supercurrent. We display their results in table \ref{tb:ansc}.

\begingroup
\small
\begin{table*}
 \begin{center}
  \begin{tabular}{|c|c|c|c|}
   \hline
    Framework & Gauge & Supercurrent   & Spin Constraint \\
	    & Invariance     & Conservation	& 	(Rarita Schwinger Constraint)	\\ \hline
    Rosenberg Method \cite{ABBOTT1} & \ding{52} & \ding{56} & \ding{52} \\ \hline
    Preregularization \cite{MACKEON} & \ding{52} (\ding{56}) & \ding{56} (\ding{52}) & \ding{52} \\ \hline
    Cutoff \cite{MACKEON}& \ding{52} (\ding{56}) & \ding{56} (\ding{52}) & \ding{52} \\ \hline
	Analytical Regularization \cite{KUMAR}&\ding{52}& \ding{56}&\ding{52} \\ \hline
    Point-Splitting \cite{INAGAKI,HAGIWARA}&\ding{52}&\ding{56} (\ding{52})&\ding{52} (\ding{56}) \\ \hline
    Dimensional Reduction \cite{MAJUMDAR,NICOLAI} & \ding{52} & \ding{56} (\ding{52})& \ding{52} (\ding{56})\\ \hline
    Dimensional Regularization \cite{CURTRIGHT,HAGIWARA}&\ding{52}&\ding{52}& \ding{56}\\ \hline
  \end{tabular} \caption{Supercurrent Anomaly in SYM $N=1$ in the Wess-Zumino gauge within different regularization schemes. Notice that the three Ward Identities cannot be satisfied simultaneously at quantum level just as the AVV triangle}  \label{tb:ansc}
 \end{center}
\end{table*}
\endgroup

\begin{figure*}
 	\begin{center}
 		\includegraphics[height=3.0cm]{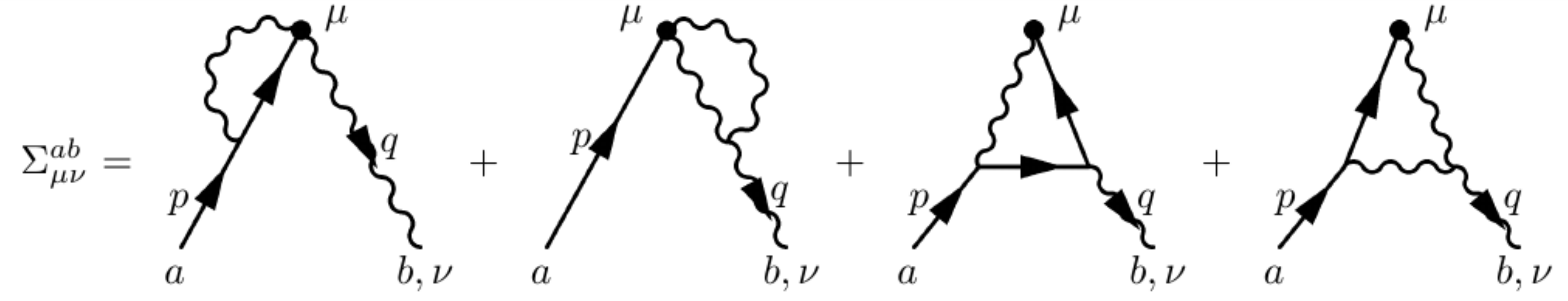}
 	\end{center}
 	\caption{Bubble and triangle graphs contributing to the process ${S_\mu} \rightarrow \psi + A_\mu$ } 
 	\label{fig:FD}
 \end{figure*}

We proceed with the calculation of the supercurrent at one loop order in a fully four dimensional setting using IReg. Feynman rules displayed in figure (\ref{fig:rf}) yield, for the diagrams depicted in figure (\ref{fig:FD}), the following superficially linearly divergent amplitude:
\begin{equation}
 \left({S}_{\mu \nu}^{ab}\right)_{1-loop} \equiv\Sigma_{\mu \nu}^{ab}=\Sigma_{\mu \nu_{A}}^{{ab}}+\Sigma_{\mu \nu_{B}}^{{ab}}+\Sigma_{\mu \nu_{C}}^{{ab}}+\Sigma_{\mu \nu_{D}}^{{ab}} \; , \label{eqs}
\end{equation}
in which
\bq
\Sigma_{\mu \nu_{A}}^{a b} &=&  \sigma_{\rho \nu}\gamma_\mu \gamma_\alpha \gamma^\rho \int_k \frac{ 4 g^2  \delta^{ab} N_A^\alpha  }{(k+s_{A})^2 (k+p+s_{A})^2 } \,\, , \nonumber \\
\Sigma_{\mu \nu_B}^{{ab}} &=& -  \sigma^{\eta \beta}\gamma_\mu \int_k\; \frac{2 g^2 \delta^{ab} N_B^{\beta \nu \eta}}{(k+s_{B})^2 (k+q+s_{B})^2}\;,  \nonumber \\
\Sigma_{\mu \nu_C}^{{ab}} &=& -  \sigma_{\alpha \zeta}\gamma_\mu \int_k \; \frac{ 4 g^2 \delta^{ab} N_{C\,\nu}^{\alpha \zeta}}{(k+s_{C})^2 (k-p+s_{C})^2 (k-q+s_{C})^2}, \nonumber \\
 \Sigma_{\mu \nu_D}^{{ab}} &=& \sigma_{\omega \eta} \gamma_\mu  \int_k\; \frac{ 4 g^2 \delta^{ab} N_{D\, \nu}^{\omega \eta }}{(k+s_{D})^2 (k-p+s_{D})^2 (k-q+s_{D})^2}\, ,\nonumber \\
 \label{AMPLIT}
\eq
with
\bq
N_A^\alpha &=& (k +p +{s_{A}})^{\alpha} \, ,\nonumber \\
N_B^{\beta \nu \eta} &=& -g_{\beta \nu}(k+2q+s_{B})_\eta + g_{\nu \eta}(-k+q-s_{B})_\beta  \nonumber \\
&+&  g_{\beta \eta}(2k + q + 2 s_{B})_\nu \, , \nonumber \\
N_{C\,\nu}^{\alpha \zeta} &=& (\s{k}-\s{q}+\s{s}_{C})\gamma_\nu(\s{k}+\s{s}_{C})\gamma^\alpha (k^\zeta-p^\zeta+s_{C}^\zeta) \, ,\nonumber \\
N_{D\, \nu}^{\omega \eta } &=& (-\s{k}+\s{p}-\s{s}_{D})\gamma_\beta (-k + q - {s_{D}})^\omega \times \nonumber \\ 
&\times& \big[(-2k+q-2 s_{D})_\nu g^{\beta \eta}+(k-2q-s_{D})^\beta \delta_{\nu}^{\eta} \nonumber \\ &+& (k+q+s_{D})^\eta \delta_{\nu}^{\beta}\big]\;.\nonumber \\
\eq
We  used $ \sigma^{\alpha \beta} \equiv \frac{1}{4} \left[\gamma^\alpha, \gamma^\beta\right]$ and,  for $SU(2)$,  $C_2(G) = 2$. Moreover
\be
 {s_{ {i}}}^\alpha \equiv m_{{i}} p^\alpha + n_{{i}} q^\alpha ,
 \label{routings}
\ee
$(i=A,\ldots, D)$ are arbitrary internal momentum routings.

 
In evaluating the integrals  displayed in (\ref{AMPLIT}),
it is important to bear in mind that, for any regularization that operates in the physical dimension,  the order in which the Clifford algebra is performed (before or after integration in internal momenta) yields different results.\footnote{The FDR scheme \cite{PITTAU}, although defined in the physical dimension, avoids this issue by appending a new set of rules to the method which ultimately results in the introduction of ``extra" integrals, absent in IReg, for instance.} In order to avoid ambiguous symmetric integrations in the physical spacetime dimension
 (and define a consistent framework with a unique relation to the quantum action principle)\cite{ADRIANOPEREZ,PEREZIRR} the algebra should be executed before integration. We illustrate this feature in appendix A.

In the amplitudes (\ref{AMPLIT}), 
we refrain to apply on-shell conditions (\ref{OSC}) at the level of the integrals which avoids the introduction of a proper regularization scheme to treat infrared divergences. Therefore, only
a fictitious mass $\mu$ is added in the propagators to regularize spurious infrared divergences that appear when  ultraviolet divergences are isolated  in terms of a BDI, namely $I_{log}(\mu^2)$. The limit $\mu\rightarrow 0$ is taken in the end of the calculation. The regularization independent relation (\ref{SR}) introduces a renormalization group scale $\lambda \ne 0$ in the BDI's in terms of which renormalization constants are defined. As the {\it{sum}} of amplitudes in equation (\ref{eqs})  contributing to the one loop correction of the supercurrent is infrared finite 
(since we refrain to apply on-shell conditions at the beginning)
 , a precise cancellation of 
spurious
infrared regulators  will take place among the various terms. 
At this stage, on-shell conditions (\ref{OSC}) can be judiciously applied to simplify the final result. We emphasize that, without a proper regularization scheme for infrared divergences, the naive application of on-shell conditions can
lead to spurious cancellations (see appendix B) as in \cite{ABBOTT1}, \cite{MACKEON}, and \cite{KUMAR}.

\section{Lorentz decomposition of the one loop correction to the supercurrent } 

Lorentz covariance allows  the following decomposition for  $\Sigma_{\mu \nu}^{ab}$
\bq
\Sigma_{\mu \nu}^{ab} 
&=& \s{q}\,\gamma_{\nu }\,\gamma_{\mu } B_0 + g_{\mu \nu } \s{q} B_1+ q_{\mu } \gamma_{\nu }  B_2 + p_{\nu } \gamma_{\mu } B_3 \nonumber \\ &+&
p_{\mu } \gamma_{\nu } B_4+  p_{\nu } q_{\mu } \s{q} B_5 +  p_{\mu } p_{\nu } \s{q} B_6\;,\label{supercc}
\eq
in which  on-shell conditions (\ref{OSC}) were taken into account (and thus the coefficients of $\g_\nu \g_\mu \s{p}$, $g_{\mu\nu}\s{p}$,  $q_\nu \g_\mu$, $q_\mu q_\nu \s{q}$ and $q_\mu q_\nu \s{p}$ are zero).

The coefficients  $B_0$, $B_1$, $B_2$, $B_3$ e $B_4$ possess regularization dependent STs (and an ultraviolet divergence in $B_0$) which is renormalizable and plays no role in the Ward identities. On the other hand $B_5$ and $B_6$ are finite and well determined (regularization independent). Notice that since the supercurrent does not couple to any field in the Lagrangian, the anomaly does not spoil the renormalizability of the theory. The Ward identities (\ref{eq:iw}) in terms of the Lorentz decomposition of the one loop correction to the supercurrent (\ref{supercc}) read
\bq
		q^\nu \Sigma^{ab}_{\mu\nu}&=&\left(B_1+B_2+p \cdot q B_5\right)q_\mu \s{q} + p \cdot q \g_\mu B_3 \nonumber \\ &+& \left(B_4+p\cdot q B_6\right)p_\mu \s{q} , \nonumber \\
		(p^\mu - q^\mu)\Sigma^{ab}_{\mu\nu}&=& \left[B_1-B_3+\left(B_5-B_6\right)p\cdot q\right]p_\nu \s{q} \nonumber  \\&+&\left(B_2-B_4\right)p\cdot q \g_\nu , \nonumber \\
		\g^\mu \Sigma^{ab}_{\mu\nu}&=& \left(B_2-B_1\right)\s{q}\g_\nu + 
	 2  (2 B_3 + B_4 + \nonumber \\ &+& p\cdot q B_6)  p_\nu .
	 \label{WIB}
\eq

In table \ref{tableWIB} we display the values assumed by the coefficients
$B_1$, \ldots, $B_4$  in terms of the coefficient $B_5$. As we shall show in the next section $B_6 = -B_5$.

\begin{table*}[]
	\centering
	\begin{tabular}{|c|c|c|c|c|c|}
		\hline
	W. I.	 & Anomaly & $B_1$  &  $B_2$ & $B_3$ &  $B_4$   \\ \hline\hline
	
	Gauge	& \ding{52} & \multirow{3}{*}{} & \multirow{3}{*}{} & \multirow{3}{*}{} & \multirow{3}{*}{} \\ \cline{1-2}
	Susy	& \ding{52} &   $- 2 B_5 (p \cdot q) $ &  $B_5 (p\cdot q)   $  &  0   &$ B_5(p \cdot q) $      \\ \cline{1-2}
	Spin-$3/2$  & $\g^\mu \Sigma^{ab}_{\mu \nu}= 3 B_5 (p\cdot q) \s{q}\g_\nu$  &                   &                   &                   &                   \\ \hline\hline
		
Gauge& \ding{52} & \multirow{3}{*}{} & \multirow{3}{*}{} & \multirow{3}{*}{} & \multirow{3}{*}{} \\ \cline{1-2}
Susy&$(p-q)^\mu \Sigma^{ab}_{\mu \nu}  = \frac{3}{2} B_5 (p\cdot q)[p_\nu \s{q} - (p\cdot q)\gamma_\nu]$  &   $- \frac{B_5}{2}(p\cdot q) $   &  $-\frac{B_5}{2} (p\cdot q) $  &   $0$    & $B_5 (p \cdot q) $   \\ \cline{1-2}
Spin-$3/2$&\ding{52} &                   &                   &                   &                   \\ \hline\hline
				
Gauge& $q^\nu \Sigma_{\mu \nu}^{ab} = -B_5 (p \cdot q) q_\mu \s{q} + B_5 (p \cdot q)^2 \gamma_\mu - 2 B_5 (p \cdot q) p_\mu \s{q}$ & \multirow{3}{*}{} & \multirow{3}{*}{} & \multirow{3}{*}{} & \multirow{3}{*}{} \\ \cline{1-2}
Susy& \ding{52}  &   $-B_5 (p\cdot q)$   &  $-B_5 (p\cdot q)$  &   $B_5 (p\cdot q)$    & $-B_5 (p\cdot q)$   \\ \cline{1-2}
Spin-$3/2$&\ding{52} &                   &                   &                   &                   \\ \hline

	\end{tabular}
	\caption{Values assumed by $B_1$, \ldots, $B_4$ coefficients in terms of the coefficient $B_5$ (momentum routing and surface term independent)   for an anomaly in each of the Ward identities. We have used the result $B_5=-B_6$.}
	\label{tableWIB}
\end{table*}

\section{On-shell Expression of  $\Sigma_{\mu \nu}^{ab}$}

In the appendix C, we exhibit details of the computation of the amplitudes which lead to the one loop correction to the supercurrent  $\Sigma_{\mu \nu}^{ab}$. Here we show the complete expression in order to discuss the relation between gauge invariance, surface terms and MRI in  Feynman diagrams. This is accomplished in a fully four dimensional framework in which, as discussed earlier,  we regularize the amplitudes taking heed of subtleties involving the non-commuting character of operations like Clifford algebra contractions and on-shell/massless limits under the integration sign of a divergent amplitude. Thus, for the coefficients appearing in equation (\ref{supercc}) we get (see appendix C): 
\bq
&& \frac{B_0}{ g^2 \delta^{ab}} = -6 I_{log} (\lambda^2) +
(-2 n_C +2  n_D +2 n_A - 13) \nu_0 \nonumber \\
&& + \frac{2}{3}b \Big[ - 3 - \pi^2 + 6 \ln \Big( \frac{-p^2}{\lambda^2}\Big)
- 9 \ln  \Big( \frac{2 p \cdot q}{\lambda^2}\Big) + \nonumber \\ && 12  \ln \Big( \frac{-q^2}{\lambda^2}\Big) + 6 \textnormal{Li}_2 \Big( 1 - \frac{2 p \cdot q}{p^2} \Big)
+ 6 \textnormal{Li}_2 \Big( 1 - \frac{2 p \cdot q}{q^2} \Big) \Big] , \nonumber \\
&& \frac{B_1}{g^2 \delta^{ab}} = 4 \Big[ (n_D - n_A -2 n_C + 3)\nu_0 + \nonumber \\ && + (n_C - n_D) \xi_0  - 2 b \Big] , \nonumber \\
&& \frac{B_2}{ g^2 \delta^{ab}} = 4 \Big[ (n_D - n_A -2 n_C - 3)\nu_0 + \nonumber \\ && + (n_C - n_D) \xi_0  +  b \Big] , \nonumber \\
&& \frac{B_3}{ g^2 \delta^{ab}} = 2 \Big[ (5 m_D + 2 m_A - 4 m_C + 2)\nu_0 + \nonumber \\ && + 2 (m_C - m_D) \xi_0  \Big] , \nonumber \\
&& \frac{B_4}{ g^2 \delta^{ab}} = 4 \Big[ ( - 2 m_A - m_C -2)\nu_0 + \nonumber \\ && + (m_C - m_D) \xi_0  + b \Big] , \nonumber \\
&& \frac{B_5}{ g^2 \delta^{ab}} =  \frac{4 b}{(p \cdot q)} = - \frac{B_6}{ g^2 \delta^{ab}}.
\label{BS}
\eq
where $\textnormal{Li}_2$ is the dilogarithm function and we made explicit the routing dependence as defined in (\ref{routings}).

Table \ref{tableWIB}  shows apparently the possibility of a ``democratic" display of the anomaly among the three Ward identities involving the supercurrent just as in the ABBJ anomaly. The $B$ coefficients involve regularization dependent terms which somewhat explain the controversial results using different frameworks exposed in table \ref{tb:ansc}. Under the light of the ABBJ anomaly example discussed earlier, one can easily verify that STs in (\ref{BS}) are always accompanied by arbitrary momentum routings and thus setting STs to zero amounts to implement MRI and consequently vector
 gauge invariance, namely 
\bq
q^\nu \Sigma^{ab}_{\mu \nu} &=& 0, \nonumber\\
(p-q)^\mu \Sigma^{ab}_{\mu \nu} &=& 0, \nonumber \\
\gamma^\mu \Sigma^{ab}_{\mu \nu} &=& \frac{3 i}{4 \pi^2} g^2 \delta^{ab} \s{q} \gamma_\nu , 
\label{RESULT}
\eq
showing that the supersymmetry Noether current remains conserved and the spin-$3/2$ constraint presents an anomaly.

\section{A digression on the results of the literature}

Although the value of the anomaly agrees in the different frameworks, there is no consensus regarding which Ward identity is anomalous at one loop level.
In \cite{ABBOTT1} the same model was discussed, and an evaluation of the anomaly in the divergence of the supercurrent was presented. Their strategy was based in imposing both gauge invariance and the spin constraint which maintains the spin-$3/2$ character of the supercurrent. The latter was obtained at the cost of factoring out the matrix product $ \sigma^{\alpha\beta}\g_\mu$ throughout the calculation in the amplitudes,
\bq
\Sigma^{ab}_{\mu\nu} &=&  \sigma^{\alpha\beta}\g_\mu \Sigma_{\alpha\beta\nu}^{ab} = \delta^{ab} \sigma^{\alpha\beta} \g_\mu [A_0(\g_\beta g_{\alpha\nu}-\g_\alpha g_{\beta\nu})\s{q} \nonumber \\ &+& 
A_1(g_{\beta \nu} p_\alpha - g_{\alpha\nu} p_\beta)+A_2(p_\alpha\g_\beta-p_\beta\g_\alpha)\g_\nu \nonumber \\
&+& A_3(q_\beta \g_\alpha-q_\alpha \g_\beta)\g_\nu+A_4(p_\alpha q_\beta  - p_\beta q_\alpha )\g_\nu \s{q} \nonumber \\ &+& A_5(q_\alpha \g_\beta-q_\beta \g_\alpha)p_\nu \s{q}+A_6(p_\beta \g_\alpha - p_\alpha \g_\beta)p_\nu\s{q}
\nonumber \\ &+& A_7(p_\beta q_\alpha-p_\alpha q_\beta)p_\nu + A_8(g_{\alpha \nu} q_\beta - g_{\beta \nu} q_\alpha)] \; ,\label{supAb}
\eq
because the identity $\gamma^\mu \sigma^{\alpha \beta } \gamma_\mu = 0$ ensures $\gamma_\mu {\cal{S}}^\mu = 0$. The coefficients $A_0, A_1, A_2, A_3$ and $A_8$ that multiply tensors of first rank, after a $shift$ in linearly divergent integrals  in $k$, contain surface terms due to the shift in the momentum integral as well as finite unambiguous terms. On the other hand $A_4, A_5, A_6$ and $A_7$ coefficients that multiply third rank tensors  are finite and unambiguous. Thus the imposition of gauge invariance, $q^\nu \Sigma_{\alpha\beta\nu}^{ab} = 0$, leads to relations among the $A_i$'s in such a way that the supercurrent anomaly is determined only by $A_4$ to $A_7$ coefficients. They applied the Rosenberg method \cite{MAJUMDAR} which assigns values for the surface terms 
so that 
Ward identities are respected, viz. gauge invariance and spin-$3/2$ constraint. The latter constraint derives from the algebraic identity $\gamma_\mu \sigma^{\alpha\beta}\g^\mu$ = 0.

The problem in maintaining the structure $\sigma^{\alpha \beta } \gamma_\mu$ factored out is that it clashes with the property of performing the Dirac algebra before integration to avoid spurious symmetric integration in the physical dimension. Whilst this approach seems a matter of choice, it is not immaterial as discussed in appendix A. More importantly,  an anomaly of the supercurrent 
 suggests   that a similar situation may develop in the context of  supergravity models (as pointed out in \cite{ABBOTT1} itself) which may pose problems due to renormalizability issues. In a subsequent work \cite{ABBOTT2}, an anomaly killing mechanism was developed at one loop level by including interactions of the Yang-Mills meson and Majorana spinors with the non-interacting scalar Wess-Zumino multiplet. Evidently we can map the expansion (\ref{supercc}) into (\ref{supAb}). For instance, the coefficients that define the spin-$3/2$ constraint in (\ref{WIB}) read
\bq
B_1 &=& B_2= 8 (A_3 - A_0 - (p\cdot q) A_4), \nonumber \\
B_3 &=& 4 [ A_1 + 2 A_2 - (2 A_6 + A_7) (p \cdot q)], \nonumber \\
B_4 &=& - 8 (A_1 + 2 A_2), \nonumber \\
B_6 &=& 8 (2A_6 + A_7),
\eq
which automatically yields zero for the spin-$3/2$ constraint in (\ref{WIB}).

In \cite{MACKEON} within a fully four dimensional approach method called Preregularization, STs are explicitly evaluated using symmetric integration (appendix B) but arbitrary momentum routings are chosen in such a way that gauge invariance or the supersymmetry Noether current is conserved at one loop level while the spin-$3/2$ constraint is maintained as their Lorentz decomposition is similar to \cite{ABBOTT1}.

Analytic regularization \cite{KUMAR} also preserves gauge invariance and the spin-$3/2$ constraint. On-shell conditions are applied before Feynman parameter integrations. Fermionic and bosonic propagators are modified in this method introducing a regularization parameter $\kappa$,
\bq
 \frac{1}{\s{k}-\s{p}} &\rightarrow& \frac{\s{k}+\s{p}}{[(k+p)^2]^{1+\kappa}} \nonumber \\
\frac{1}{(k-q)^2} &\rightarrow& \frac{1}{[(k-q)^2]^{1+\kappa}}  . 
\eq 
In order to preserve gauge invariance in analytic regularization, in (\ref{eqs}) either the bosonic ($\Sigma_{\mu \nu \, A}^{ab}$ and $\Sigma_{\mu \nu \, B}^{ab}$) or the fermionic ($\Sigma_{\mu \nu \, C}^{ab}$ and $\Sigma_{\mu \nu \, D}^{ab}$) propagators should be modified. Using Point-Splitting regularization, in \cite{INAGAKI} was shown an anomaly for  spinor current which however could be removed by a redefinition of the gauge invariant conserved current keeping the spin-$3/2$ constraint satisfied. 

Hagiwara and collaborators \cite{HAGIWARA}, on the other hand, employed dimensional , Pauli-Villars  and Point-splitting regularizations to show that there exists an anomaly in the superconformal current but not in the divergence of the supercurrent. They emphasize the importance of picking up a definite regularization in studying the anomaly of the supercurrent. Otherwise, due to its specific structure, all the contributions to the one loop correction would have the form  $\sigma^{\alpha \beta} \gamma_\mu \Sigma_{\alpha\beta\nu}$ which naturally grants  privilege to the spin-$3/2$ constraint to be satisfied and consequently $\partial^\mu {\cal{S}}_\mu \ne 0$.  In reference \cite{SMAILAGIC}, it is argued that it is not a matter of simply redefining loop momenta that would shift the anomaly from the supersymmetric current into the spin-$3/2$ constraint as this would have important implications on the multiplet structure of currents and anomalies. 

In \cite{MEHTA}, the supercurrent and superconformal anomalies were evaluated for off-shell $N=1$ supersymmetric Yang-Mills theory within the Fujikawa method and the heat kernel regularization scheme. They obtain no one-loop supercurrent anomaly and a superconformal anomaly that agrees with our calculation.

Finally, dimensional methods were discussed in \cite{CURTRIGHT,MAJUMDAR,HAGIWARA,NICOLAI}. In \cite{CURTRIGHT}, conventional dimensional regularization was employed, setting the anomaly at the superconformal sector, a result that was later confirmed by \cite{HAGIWARA}. In \cite{NICOLAI}, dimensional reduction was used, with the same outcome. As discussed in this reference, for the calculation at hand at least at one-loop order, the two schemes are equivalent. The reason is that $\epsilon$-scalars, which need to be introduced in the dimensional reduction scheme to avoid a mismatch between fermionic and bosonic degrees of freedom, do not contribute to the Ward identities studied here. In \cite{MAJUMDAR}, instead of embedding the theory in a quasi 4-dimensional space (which will enforce the introduction of $\epsilon$-scalars, for instance), the supercurrent was defined in strictly 4-dimensions. Therefore, the identity $\gamma^\mu \sigma^{\alpha \beta } \gamma_\mu = 0$ is satisfied by construction which leads to $\gamma_\mu {\cal{S}}^\mu = 0$. Since gauge symmetry still holds in this scheme, the only available Ward identity to be violated will be $\partial^\mu {\cal{S}}_\mu$, which is the result found by the authors.

These results are all summarized in table \ref{tb:ansc}.

\section{Discussion and Conclusions}

It is well known that in dimensional specific models which includes chiral, supersymmetric and topological field theories, Feynman diagram calculations cannot be handled in standard dimensional regularization. Despite some generalizations in dimensional methods to tackle the algebra of parity violating objects such as the $\gamma_5$-matrix   constructed in \cite{MAISON}, issues related to the possibility of spurious anomalies and bundersome Ward identities to be imposed order by order on the Green's functions make it worthwhile seeking for a framework that fully operates in the physical dimension. Yet IReg adequately addresses the problem of evaluation of undetermined regularization dependent surface terms related to momentum routing in Feynman diagrams to all loop orders, care must be exercised in the formal treatment of Lorentz tensors and $\gamma_5$ matrices as discussed in \cite{ADRIANOPEREZ} even in non-dimensional methods \cite{Viglioni2016,JOILSON}. In the particular instance of the supercurrent anomaly, we have seen that Clifford algebra does not commute with integration in the loop momenta. Novel schemes that do not rely on dimensional continuation in the space time dimension have been proposed and developed. The motivation for this progress has been to broaden the conceptual basis as well as to enable 
efficient, automated analytical and numerical calculational methods to test beyond the standard model theories within precision observables.

Relying on a fully four dimensional approach in which regularization dependent terms are left to be fixed on symmetry grounds, we have calculated the supercurrent anomaly of the zero mass Yang-Mills multiplet interacting with a single massless Majorana spin $1/2$ field transforming in the adjoint representation of $SU(2)$ for simplicity.  Momentum routing invariance, an obviously desired property of Feynman diagram calculations, is known to be connected to gauge invariance as discussed in the introduction. We have seen that setting surface terms to zero automatically implements momentum routing invariance in the one loop correction to the supercurrent. The quadridivergence of the supercurrent as well as gauge symmetry remains conserved at quantum level, whereas the superconformal invariance translated by the spin-$3/2$ constraint is broken as seen in  equation (\ref{RESULT}). 

Unlike the ABBJ anomaly in which the anomaly shifts between the Ward identities as shown in equations (\ref{WI22}), the violation of one of the three Ward identities as shown in table \ref{tableWIB} is not effected by a simple choice of STs, displacing the anomaly from one Ward identity to another. This suggests that the ``democracy" in the perturbative calculation of the anomaly as seen in ABBJ anomaly \cite{JACKIWFU} does not occur.  For example, for an anomalous supercurrent conservation,
\be
(p-q)^\mu \Sigma ^{ab}_{\mu \nu} = \frac{3ig^2}{8 \pi^2} \delta^{ab} (p_\nu \s{q} - (p \cdot q) \gamma_\nu)
\ee
and spin-$3/2$ constraint and gauge invariance satisfied one must have both $\nu_0 = b/2$, $\xi_0 = 5 b/6$ and $5 m_D + 6 m_A - 2 m_C = -6$, $2 n_D + 3 n_A + n_C = 0$. On the other hand,  an anomalous gauge Ward identity,
\be
q^\nu \Sigma ^{ab}_{\mu \nu} = \frac{ig^2}{4 \pi^2} \delta^{ab} [(p \cdot q)\gamma_\mu  - 2 p_\mu \s{q} - q_\mu \s{q}]
\ee
and supercurrent and spin-$3/2$ conservation fulfilled is obtained only if  $\nu_0 = b/2$, $\xi_0 = 5 b/6$ and $5 m_D + 6 m_A - 2 m_C = 6$, $2 n_D + 3 n_A + n_C = 3$.

Our results in IReg agree with those obtained in
dimensional regularization and dimensional reduction (in the case in which the theory is embedded in a quasi 4-dimensional space as customary). 
Although at one-loop order there is
no difference between these two dimensional schemes for the Ward identities here studied,
 we believe that IReg would reproduce the results of dimensional reduction at higher loop (in this case, it is expected that the dimensional schemes differ by finite terms, due to the inclusion of $\epsilon$-scalars contributions). This fact has recently been observed in the case of non-supersymmetric theories \cite{ZURICH,ADRIANOPEREZ}. Finally, it should be noticed that, as reported in \cite{ADRIANOPEREZ}, the correlation between IReg and dimensional schemes is only possible if subtleties involving the Clifford algebra are dealt with properly.
 By doing so  one ultimately obtains a consistent framework for Feynman diagram calculations in the physical dimensions, in the sense that all regularization dependent terms that appear are directly connected with MRI and gauge symmetry as in \cite{Adriano2012,Viglioni2016}.

\section*{Acknowledgements}
M. S. acknowledges research Grant No. 303482/2017-6 784 from CNPq-Brazil and financial
support by FAPESP. B.H. acknowledges partial support from the FCT (Portugal) project UID/FIS/04564/2016. A.C.\ acknowledges financial support from CAPES (Coordena\c{c}\~{a}o de Aperfei\c{c}oamento de
Pessoal de N\'{i}vel Superior), Brazil. B.H. and A.C. would like to acknowledge networking support by the COST Action CA16201.

\section*{Appendix A}

Consider the following  divergent piece of an amplitude in four dimensions,
\be
\gamma_{\mu}\gamma_{\nu} \int_k \frac{k^\mu k^\nu}{(k^2-m^2)^3} \equiv \gamma_{\mu}\gamma_{\nu} I_{log}^{\mu \nu} (m^2).
\label{Ilogmunu}
\ee
Performing the Dirac algebra and using that $\slashed{k}^2 = k^2$ gives
\be
\gamma_{\mu}\gamma_{\nu} I_{log}^{\mu \nu} (m^2) = I_{log}(m^2) + \int_k\frac{m^2}{(k^2-m^2)^3},
\label{before}
\ee
where the second term on the RHS is finite and regularization independent. On the other hand, using (\ref{dif1}) enables us to write
\bq
\gamma_{\mu}\gamma_{\nu} I_{log}^{\mu \nu} (m^2) &=& \gamma_{\mu}\gamma_{\nu}\Big[ \frac{g_{\mu \nu}}{4} I_{log}(m^2) + \frac{g_{\mu \nu}}{4} \upsilon_0\Big]\nonumber \\
&=&  I_{log}(m^2) + \upsilon_0 ,
\label{after}
\eq
which is the result performing the Dirac algebra after manipulating the integrand (and eventually performing the integration). Recall that $\upsilon_0$ is a regularization dependent surface term. Such an ambiguity is present in any strictly four dimensional regularization
(as long as one refrains to append a set of rules to circumvent this problem as done in the FDR scheme)
  and a prescription must be adopted
in order to avoid spurious symmetry breaking. As a matter of fact such an ambiguity is related to {\it{ symmetric integration}} which at one loop level means to set
\be
\int_k  k^{\mu_1} k^{\mu_2} .. k^{\mu_{2n}} f(k^2) = 
\int_k \frac{g^{\{\mu_1 \mu_2} .. g^{\mu_{2n-1} \mu_{2n} \}} k^{2n}}{(2n)!}  f(k^2),
\ee
the curly brackets standing for symmetrization on Lorentz indices, which is well known to be a forbidden operation for divergent integrals in the physical dimension \cite{PEREZIRR} while it is allowed within dimensional regularization. In order to avoid symmetrical integration one should perform the Dirac algebra (contractions, traces) before manipulating the amplitude integrand. Indeed, within dimensional regularization, performing the $\gamma$-matrices contraction before integration, 
\bq
\gamma_{\mu}\gamma_{\nu} I_{log}^{\mu \nu} (m^2)\Big|_{DR}^{before} &=& \gamma_\mu \gamma_\nu \frac{g_{\mu \nu}}{d} \int_{k}^{d} \frac{k^2}{(k^2-m^2)^3}\nonumber \\
&=& \frac{d}{d}\, b \, \big(\frac{4 \pi}{m^2}\big)^\epsilon (1-\epsilon/2)\Gamma(\epsilon),
\label{rdbefore}
\eq
whereas using that
\be
I_{log}^{\mu \nu} (m^2)\Big|_{DR} = \frac{b}{4}\big(\frac{4 \pi}{m^2}\big)^\epsilon \Gamma(\epsilon) g^{\mu \nu},
\ee
gives
\be
\gamma_{\mu}\gamma_{\nu} I_{log}^{\mu \nu} (m^2)\Big|_{DR}^{after} =  \frac{b}{4}\big(\frac{4 \pi}{m^2}\big)^\epsilon \Gamma(\epsilon) (4-2 \epsilon).
\label{rdafter}
\ee
with $b$ defined as in (\ref{b}), which is just the same result as  (\ref{rdbefore}), namely
\be
\frac{b}{\epsilon} - b \, \gamma_E + b \ln \big(\frac{4 \pi}{m^2}\big) - \frac{b}{2}.
\label{DiracDR}
\ee
On the other hand, while equation (\ref{before}) reproduces the results of DR displayed in (\ref{DiracDR}), performing the Dirac algebra after integration as in (\ref{after}) yields, in DR 
\be
\frac{b}{\epsilon} - b \,  \gamma_E + b \ln \big(\frac{4 \pi}{m^2}\big),
\ee
that differs by a term $-b/2$ from (\ref{DiracDR}). Such a term can spuriously break  symmetries in the underlying theory. Thus, should we use nondimensional methods to tackle divergent Feynman amplitudes of theories that are sensitive to dimensional continuation on the spacetime dimension, care must be exercised with both the Clifford algebra which does not commute with integration and symmetric integration \cite{ADRIANOPEREZ}.
\section*{Appendix B}
Consider the UV divergent and IR finite integral
(considering that on-shell conditions such as $p^2=0$ are not immediately applied),
\begin{equation}
\int_k \frac{1}{k^2(k+p)^2}\; . 
\end{equation}
We evaluate it by introducing an infrared regulator $\mu$ in intermediate steps and taking the limit $\mu \rightarrow 0$. The regularization independent relation (\ref{SR}) introduces the renormalization scale $\lambda^2 \ne 0$ 
as below
\bq
\int_k \frac{1}{k^2(k+p)^2} &=& \lim_{\mu\rightarrow 0} \int_k \frac{1}{(k^2-\mu^2)[(k+p)^2-\mu^2]} \nonumber \\ 
&=& I_{log}(\lambda^2)
-b   \ln\left(-\frac{p^2}{\lambda^2}\right) + 2 b. \nonumber
\eq
However applying on-shell conditions from the start yields
\bq
&&\lim_{\mu\rightarrow0} \int_k \frac{1}{(k^2-\mu^2)[(k+p)^2-\mu^2]} = \nonumber \\
&& \lim_{\mu \rightarrow0} \left[\int_k \frac{1}{(k^2-\mu^2)^2}-2 p_\alpha \int_k \frac{k^\alpha}{(k^2-\mu^2)^2[(k^2-2k\cdot p-m^2)]}\right]  \nonumber \\
&&\overset{p^2\downarrow0}{=}\lim_{\mu\rightarrow 0}\left[\int_k \frac{1}{(k^2-\mu^2)^2}-4p_\alpha\int_0^1{d}z \, z\int_k \frac{k^\alpha+p^\alpha(z-1)}{[k^2-\mu^2]^3}\right] 
\nonumber \\
&&= I_{log}(\lambda^2) - b \lim_{\mu \rightarrow0} \ln\left(\frac{\mu^2}{\lambda^2}\right). \nonumber
\eq
Notice that the two results differ not only by logarithmically infrared divergent terms (setting $p^2\rightarrow 0$ in the first case and $\mu^2\rightarrow 0 $ in the second), but also by a constant $2 b$. Therefore, to use the latter, a proper infrared regularization scheme is needed, since without it spurious cancellations in the case of the full one-loop supercurrent calculation may take place.

\section*{Appendix C}
The coefficients $B_i$, $i= 0, \ldots, 6$ in the Lorentz decomposition of the one loop correction to the supercurrent (\ref{supercc})
receive contributions from each of the amplitudes in equation (\ref{AMPLIT}) that represent the diagrams of figure \ref{fig:FD}, namely
\be
B_i = B_i^A +  B_i^B +  B_i^C +  B_i^D, \,\,\, i= 0, \ldots, 6. 
\ee
Here we display each contribution separately as well as some useful results of integrals cast in IReg.
\bq
-2 B_0^A &=& B_1^A =  B_2^A =-4g^2\delta^{ab} n_A \nu_0, \nonumber \\
B_3^A &=& - \frac{1}{2} B_4^A = 2 g^2\delta^{ab} \big[I_{log}(\lambda^2) +(2m_A+1)\nu_0 \nonumber \\
&-& b \ln\big(- \frac{p^2}{\lambda^2}\big) + 2b \big],\nonumber \\
B_5^A &=& B_6^A = 0.
\eq
\bq
B_0^B &=& -3g^2\delta^{ab} \big[I_{log}(\lambda^2) 
- b \ln\Big(- \frac{q^2}{\lambda^2}\Big) + 2b \big],\nonumber \\
B_1^B &=& B_2^B =B_3^B = B_4^B = B_5^B = B_6^B = 0.
\eq

\bq
B_0^C &=& g^2 \delta^{ab} \big[ -I_{log}(\lambda^2) - (2 n_C+5)\nu_0 + \nonumber \\ &+&b \ln \Big( \frac{2 (p\cdot q)}{\lambda^2}\Big) + 2b    \big],
\eq
\bq
B_1^C &=& \frac{2g^2}{9} \delta^{ab} \big[ - 3 I_{log}(\lambda^2) + 9 (3- 4 n_C)\nu_0 + \nonumber \\ &+& 6 (3 n_C- 1)\xi_0  + 3 b \ln \Big( \frac{2 (p\cdot q)}{\lambda^2}\Big) - 20 b    \big],
\eq
\bq
B_2^C &=& \frac{2g^2}{9} \delta^{ab} \big[ - 3 I_{log}(\lambda^2) - 9 (1 + 4 n_C)\nu_0 + \nonumber \\ &+& 6 (3 n_C- 1)\xi_0  + 3 b \ln \Big( \frac{2 (p\cdot q)}{\lambda^2}\Big) - 2 b    \big],
\eq
\bq
B_3^C &=& \frac{2g^2}{9} \delta^{ab} \big[ - 3 I_{log}(\lambda^2) + 9 (1- 4 m_C)\nu_0 + \nonumber \\ &+& 6 (3 m_C- 1)\xi_0  + 3 b \ln \Big( \frac{2 (p\cdot q)}{\lambda^2}\Big) - 2 b    \big],
\eq
\bq
B_4^C &=& \frac{4g^2}{9} \delta^{ab} \big[  3 I_{log}(\lambda^2) -9 m_C \nu_0 + \nonumber \\ &+& 3 (3 m_C- 1)\xi_0  - 3 b \ln \Big( \frac{2 (p\cdot q)}{\lambda^2}\Big) + 8 b    \big],
\eq
\be
 B_5^C = -2 B_6^C = -\frac{8g^2\delta^{ab}}{3}\frac{b}{(p\cdot q)}.
\ee

\bq
B_0^D &=& g^2 \delta^{ab} \Big[ - 2 I_{log}(\lambda^2) + 2 ( n_D -4 )\nu_0 \nonumber \\ &+& 2b -\frac{2b\pi^2}{3} + 4 b \ln \Big( -\frac{p^2}{\lambda^2}\Big) -7b \ln \Big( \frac{2 (p\cdot q)}{\lambda^2}\Big)  \nonumber \\ &+& 5b \ln \Big(- \frac{q^2}{\lambda^2}\Big) + 4b \textnormal{Li}_2 \Big( 1 - \frac{2 (p \cdot q)}{p^2} \Big)  \nonumber \\ &+&  4b \textnormal{Li}_2 \Big( 1 - \frac{2 (p \cdot q)}{q^2} \Big)\Big],
\eq
\bq
B_1^D &=& \frac{2g^2}{9} \delta^{ab} \big[  3 I_{log}(\lambda^2) + 9 (3 + 2 n_D)\nu_0 + \nonumber \\ &+& 6 (1 - 3 n_D)\xi_0  - 3 b \ln \Big( \frac{2 (p\cdot q)}{\lambda^2}\Big) - 16 b    \big],
\eq
\bq
B_2^D &=& \frac{2g^2}{9} \delta^{ab} \big[  3 I_{log}(\lambda^2) + 9 (-5 + 2 n_D)\nu_0 + \nonumber \\ &+& 6 (1 - 3 n_D)\xi_0  - 3 b \ln \Big( \frac{2 (p\cdot q)}{\lambda^2}\Big) + 20 b    \big],
\eq
\bq
B_3^D &=& \frac{2g^2}{9} \delta^{ab} \Big[ - 6 I_{log}(\lambda^2) + 45 m_D \, \nu_0 \nonumber \\ &+& 6 (1- 3 m_D)\xi_0  - 16 b  + 9 b \ln \Big( -\frac{p^2}{\lambda^2}\Big) \nonumber \\ &-& 3 b \ln \Big( \frac{2 (p\cdot q)}{\lambda^2}\Big) \Big],
\eq
\bq
B_4^D &=& \frac{4ig^2}{9} \delta^{ab} \Big[  6 I_{log}(\lambda^2) - 9 \nu_0 \nonumber \\ &+& 3 (1- 3 m_D)\xi_0  + 19 b  - 9 b \ln \Big( -\frac{p^2}{\lambda^2}\Big) \nonumber \\ &+& 3 b \ln \Big( \frac{2 (p\cdot q)}{\lambda^2}\Big) \Big],
\eq
\be
-2B_5^D =  B_6^D =- \frac{8g^2\delta^{ab}}{3} \frac{b}{(p \cdot q)}.
\ee
The following integrals are useful (the on-shell limits ($p^2 \rightarrow 0, q^2 \rightarrow 0$) have already been judiciously taken):
\be
 \int_k \frac{1}{k^2(k+p)^2}= {I}_{log}(\lambda^2)+2b + b \ln\left(-\frac{\lambda^2}{p^2}\right),
 \ee
\bq
 \int_k \frac{k^\alpha}{k^2(k+p)^2}&=&\frac{p^\alpha}{2} \bigg[-{I}_{log}(\lambda^2)+ \nu_0 - 2b \nonumber \\
	&-&b \ln\left(-\frac{\lambda^2}{p^2}\right)\bigg],
\eq	
\bq
&& \int_k \frac{1}{k^2(k-p)^2(k-q)^2}= \frac{b}{2 p \cdot q} \bigg[-4 \ln^2 2-
\nonumber \\&& -4 \ln2 \ln\left(-\frac{p\cdot q}{q^2}\right)- \ln^2\left(-\frac{p\cdot q}{q^2} \right)+ \frac{\pi^2}{3}\Big],
\eq	

\bq
 \int_k \frac{k^\alpha}{k^2(k-p)^2(k-q)^2} &=&
\frac{b}{2 p\cdot q}\bigg[-p^\alpha \ln\left(-\frac{p^2}{2p\cdot q}\right) \nonumber \\ &+& q^\alpha\ln\left(-\frac{q^2}{2p\cdot q}\right)\bigg],
\eq	

\bq
\int_k \frac{k^2}{k^2(k-p)^2(k-q)^2} &=& {I}_{log}(\lambda^2)+2b \nonumber \\ 
&-& b \ln\left(\frac{2 p\cdot q}{\lambda^2}\right), 
\eq	
\bq
&& \int_k \frac{k^\alpha k^\beta}{k^2(k-p)^2(k-q)^2} = \frac{g^{\alpha \beta}}{4} \bigg[{I}_{log}(\lambda^2)-\nu_0\nonumber \\
&& +b \left[3-\ln\left(\frac{2p\cdot q }{\lambda^2}\right)\right]\bigg] +\frac{b}{4 p\cdot q} \bigg[ p^\alpha p^\beta \ln\left(-\frac{p^2}{2p\cdot q}\right) \nonumber \\ && -q^\alpha q^\beta \ln\left(-\frac{q^2}{2p\cdot q}\right)+\left(p^\alpha q^\beta+p^\beta q^\alpha\right) \bigg],
\eq	
\bq
&&\int_k \frac{k^\alpha k^\beta k^\eta}{k^2(k-p)^2(k-q)^2} = \frac{1}{36}p^{\{\eta } g^{\alpha \beta \}} \bigg[3 {I}_{\log }\left(\lambda^2\right) \nonumber \\
&& - 3  \xi_0 +b \left(8+3 \ln\left(\frac{\lambda ^2}{2 p\cdot q}\right)\right)\bigg] \nonumber \\ &&
+\frac{1}{36} q^{\{\eta } g^{\alpha \beta \}} \bigg[3 {I}_{\log }\left(\lambda ^2\right) - 3 \xi_0 + b \bigg(8 + 3 \ln \left(\frac{\lambda ^2}{2 p\cdot q}\right)\bigg)\bigg] \nonumber \\ && -\frac{b}{12 p\cdot q}\left(p^{\{\alpha } p^{\beta }   q^{\eta \}}+p^{\{\alpha } q^{\beta }   q^{\eta \}}\right) \nonumber \\
&& +\frac{b p^{\alpha } p^{\beta } p^{\eta }}{6 p\cdot q} \ln \left(-\frac{p^2}{2 p\cdot q}\right)+\frac{b q^{\alpha } q^{\beta } q^{\eta }}{6 p\cdot q}\ln \left(-\frac{q^2}{2 p\cdot q}\right),
\eq	
\bq 
&& \int_k \frac{k^2 k^\alpha}{k^2(k-p)^2(k-q)^2} =\frac{(p+q)^{\alpha }}{2} \bigg[{I}_{\log }\left(\lambda^2\right) \nonumber \\
&& - \nu_0 +b\left(2- \ln \left(\frac{2 p\cdot q}{\lambda^2}\right) \right)\bigg],
\eq
where
\bq
p^{\{\eta } g^{\alpha \beta \}} &=& p^{\eta } g^{\alpha \beta }+p^{\beta } g^{\alpha \eta }+p^{\alpha} g^{\beta \eta } \nonumber \\
p^{\{\alpha } p^{\beta }   q^{\eta \}} &=& p^{\beta } p^{\eta }   q^{\alpha }+p^{\alpha } p^{\eta } q^{\beta }+p^{\alpha } p^{\beta }   q^{\eta }.\nonumber
\eq

\section*{Appendix D}

We perform the computation of the finite part of the triangle diagram, $\widetilde{T}_{\mu\nu\alpha}$. Since it does not
depend on the routing, we  choose $k_1=0$, $k_2=q$ e $k_3=-p$ to get :

\bq
&& T_{\mu\nu\alpha}=-i\int_k Tr\left[\gamma_{\mu}\frac{i}{\slashed{k}-m}\gamma_{\nu}\frac{i}{\slashed{k}+\slashed
	{q}-m}\gamma_{\alpha}\gamma_5 \frac{i}{\slashed{k}-\slashed{p}-m}\right]+ \nonumber \\
&& + (\mu \leftrightarrow \nu, p\leftrightarrow q)= -8i\upsilon_0 \epsilon_{\mu \nu \alpha \beta}(q-p)^{\beta}+\widetilde{T}_{\mu\nu\alpha},
\eq
in which
\bq
\widetilde{T}_{\mu\nu\alpha}&=& 4i b\{\epsilon_{\alpha\mu\nu\lambda}q^{\lambda}(p^2\xi_{01}(p,q)-q^2
\xi_{10}(p,q)) \nonumber \\ &+& \epsilon_{
	\alpha\mu\nu\lambda}q^{\lambda}(1+2m^2\xi_{00}(p,q)) 
 4\epsilon_{\alpha\nu\lambda\tau}p^{\lambda}q^{\tau}[(\xi_{01}(p,q) \nonumber \\ &-& \xi_{02}(p,q))p_{\mu}+\xi_{11}(p,q)q_{\mu}] 
(\mu \leftrightarrow \nu, p\leftrightarrow q)\}.
\label{Tfinito}
\eq
The functions $\xi_{nm}(p,q)$ are defined as
\be
\xi_{nm}(p,q)=\int^1_0 dz\int^{1-z}_0 dy \frac{z^n y^m}{Q(y,z)},\\
\ee
with
\be
Q(y,z)=[p^2 y(1-y)+q^2 z(1-z)-2(p\cdot q)yz-m^2]
\ee
obeying the property $\xi_{nm}(p,q)=\xi_{mn}(q,p)$.

Those integrals satisfy the following relations which we have also used in the derivation of equation (\ref{Tfinito})
\bq
&&q^2 \xi_{11}(p,q)-(p\cdot q)\xi_{02}(p,q)=\frac{1}{2}\big[ -\frac{1}{2}Z_0((p+q)^2,m^2)\nonumber \\ &+&
\frac{1}{2}Z_0(p^2,m^2)+q^2 \xi_{01}(p,q)
\big],\label{f1}\\
&&p^2 \xi_{11}(p,q)-(p\cdot q)\xi_{20}(p,q)=\frac{1}{2}\big[ -\frac{1}{2}Z_0((p+q)^2,m^2)\nonumber \\ &+&\frac{1}{2}Z_0(q^2,m^2)+p^2 \xi_{10}(p,q)
\big],\label{f2}\\
&&q^2 \xi_{10}(p,q)-(p\cdot q)\xi_{01}(p,q)=\frac{1}{2}[ -Z_0((p+q)^2,m^2)\nonumber \\ &+&Z_0(p^2,m^2)+q^2 \xi_{00}(p,q)],\label{f3}\\
&&p^2 \xi_{01}(p,q)-(p\cdot q)\xi_{10}(p,q)=\frac{1}{2}[ -Z_0((p+q)^2,m^2)\nonumber \\ &+&Z_0(q^2,m^2)+p^2 \xi_{00}(p,q)],\label{f4}\\
&&q^2 \xi_{20}(p,q)-(p\cdot q)\xi_{11}(p,q)=\frac{1}{2}\big[-\big(\frac{1}{2}+m^2\xi_{00}(p,q)\big) \nonumber \\ &+&\frac{1}{2}p^2\xi_{01}(p,q)+\frac
{3}{2}q^2\xi_{10}(p,q)\big],\label{f5}\\
&&p^2 \xi_{02}(p,q)-(p\cdot q)\xi_{11}(p,q)=\frac{1}{2}\big[-\big(\frac{1}{2}+m^2\xi_{00}(p,q)\big)\nonumber \\ &+&\frac{1}{2}q^2\xi_{10}(p,q)+\frac
{3}{2}p^2\xi_{01}(p,q)\big],\label{f6}
\eq
where $Z_k(p^2,m^2)$ is defined as
\be
Z_k(p^2,m^2)=\int^1_0 dz z^k \ln \frac{m^2-p^2 z(1-z)}{m^2}.
\ee
The derivation of the relations(\ref{f1})-(\ref{f6}) are easily performed by integration by parts. A complete review on the evaluation of one loop n-point functions in a four dimensional set up can be found in \cite{DALLABONA2}.


\begin{thebibliography}{99}
	\bibitem{ADLER} S. L. Adler,  ``{\it{Anomalies to all orders}}" in 50 years of Yang-Mills theory, (2005) World Scientific, doi: 10.1142/9789812567147-0009
	\bibitem{ABBJ} S. Adler, Phys. Rev. 177 (1969) 2426 ; J. Bell and R. Jackiw, Nuovo Cim. 60A (1969) 47; S. Adler, Phys. Rev. 182 (1969) 1517; W. A. Bardeen, Phys. Rev. 184 (1969) 1848.
	
	\bibitem{EBERT} D. Ebert and H. Reinhardt, Nuc. Phys. B271 (1986) 188.
	
	\bibitem{BERN} W. Bernreuther, Lect. Notes Phys. 591 (2002) 237.
	\bibitem{HARVEY} J. A. Harvey, ``{\it{Lectures on Anomalies}}", TASI 2003, hep-th/0509097v1.
	\bibitem{BERTLMANN} R. A. Bertlmann, ``{\it{ Anomalies in Quantum Field Theory}}", Oxford University Press, (2000).
	\bibitem{JJ} I. Jack and D. R. T. Jones,  in ``{\it{ Perspectives on Supersymmetry}}", World Scientific, Ed. G. Kane, (1997).
	\bibitem{KRAUSAN} E. Kraus,  Phys. Rev. D65 (2002) 105003.
	\bibitem{KRAUSSU} E. Kraus, Nucl. Phys. B620 (2002) 55, hep-th/0107239.
	\bibitem{SCHWARZ} J. H. Schwarz, Phys. Lett. B371 (1996) 223 ; M. B. Green, J. H. Schwarz and P. C. West, Nuc. Phys. B254 (1985) 327.
	\bibitem{KRAUS} W. Hollik, E. Kraus and D. Stockinger, Eur. Phys. J. C11 (1999) 365.
	
	\bibitem{BECCHI} C. Becchi, A. Ronet and R. Stora, Phys. Lett. B52 (1974) 344 ; Commun. Math. Phys. 42 (1975) 127 ; Ann. of Phys. 98 (1976) 287.
	\bibitem{WHITE} P. L. White, Class. Q. Grav. 9 (1992) 1663.
	\bibitem{PIGUET} N. Maggiore, O. Piguet and S. Wolf, Nuc. Phys. B458 (1996) 403.
	\bibitem{DR} G. 't Hooft and M. Veltman, Nucl. Phys. B44 (1972) 189.
	\bibitem{SIEGEL} W. Siegel, Phys. Lett. B84 (1979) 193 ; Phys. Lett. B94 (1980) 37.
 	\bibitem{STOCKINGERQAP} D. Stockinger, JHEP 0503 (2005) 076.	
	\bibitem{STOCKINGERNPB} A. Signer and D. Stockinger, Nucl. Phys. B808 (2009) 88.
	\bibitem{MAISON}  P. Breitenlohner and D. Maison in ``{\it{Renormalization of Supersymmetric Yang-Mills theories}}", in Proceedings : Supersymmetry and its Applications, p. 309, Cambridge (1985) 
	
	
	
	\bibitem{JUNG} D.-W. Jung and J. Y. Lee, JHEP 0903 (2009) 123.
	
	\bibitem{FZSS} S. Ferrara and B. Zumino, Nuc. Phys. B79 (1974) 413; A. Salam and J. Strathdee, Phys. Lett. 51B (1974) 353.
	
	\bibitem{AGUILA} F. Del Aguila, A. Culatti, R. Munoz Tapia and M. Perez-Victoria, Nucl. Phys. B537 (1999) 561.
	
	\bibitem{ORIMAR} Orimar A. Battistel, PhD thesis, Universidade Federal de Minas Gerais, ICEx, Departamento de F\'{\i}sica, 2002.
	\bibitem{IR} O. A. Battistel, A. L. Mota and M. C. Nemes, Mod. Phys. Lett. A13 (1998) 1597.
	
	\bibitem{Viglioni2016} A. C. D. Viglioni, A. L. Cherchiglia, A. R. Vieira, B. Hiller and M. Sampaio, Phys. Rev. D94  (2016) 065023.
	\bibitem{Vieira} A. R. Vieira, A. L. Cherchiglia and M. Sampaio, Phys. Rev. D93 (2016) 025029 
	\bibitem{BaetaScarpelli2001}  A. P. Baeta Scarpelli, M. Sampaio and M. C. Nemes,  Phys. Rev.  D63 (2001) 046004.
	\bibitem{Cherchiglia2010} A. L. Cherchiglia, M. Sampaio and M. C. Nemes,  Int. J. Mod. Phys. A26  (2011) 2591.
	\bibitem{Adriano2012} L. C. Ferreira, A. L. Cherchiglia, B. Hiller, M. Sampaio and M. C. Nemes, Phys. Rev. D86 (2012) 025016.
	\bibitem{Cherchiglia2014} A.L. Cherchiglia, A.R. Vieira, B. Hiller, A.P. Baêta Scarpelli, M. Sampaio, Annals Phys. 351 (2014) 751.
	\bibitem{Pontes} C. R. Pontes, A.P. Ba\^eta Scarpelli, M. Sampaio and M. C. Nemes, J. Phys. G34 (2007) 2215.
	\bibitem{Dias} E. W. Dias,  A.P. Ba\^eta Scarpelli, L.C.T. Brito, M. Sampaio and M. C. Nemes Eur. Phys. J. C55 (2008) 667.
	\bibitem{Sampaio2002}  M. Sampaio, A. P. Ba\^eta Scarpelli, B. Hiller, A. Brizola, M. C. Nemes and S. Gobira, Phys. Rev. D65 (2002) 125023.
	\bibitem{BaetaScarpelli20012}  A. P. Baeta Scarpelli, M. Sampaio, B. Hiller and M. C. Nemes, Phys. Rev. D 64 (2001) 046013.
	\bibitem{Souza2005}  L. A. M. Souza, M. Sampaio and M. C. Nemes,  Phys. Lett. B632 (2006) 717.
	\bibitem{IRI} H. Fargnoli, A. Paulo Ba\^eta Scarpelli, B. Hiller, M. Sampaio and A. Osipov, Mod. Phys. Lett. A26 (2011) 289.   
	\bibitem{Ottoni2006}  J. E. Ottoni, A. P. Baeta Scarpelli, M. Sampaio and M. C. Nemes, Phys. Lett. B642 (2006) 253.
	
	
	\bibitem{FARGNOLISUSY} H. Fargnoli, B. Hiller, Marcos Sampaio and M. C. Nemes, Eur. Phys. J. C71 (2011) 1633.
	\bibitem{CHERCHIGLIAEPJC2016} A. Cherchiglia, Marcos Sampaio, B. Hiller and A. P. B. Scarpelli, Eur. Phys. J. C76 (2016) 47.
	
	
	\bibitem{Scarpelli2008}  A. P. Ba\^eta Scarpelli, M. Sampaio, M. C. Nemes and B. Hiller, Eur.  Phys.  J. C56 (2008) 571.
	\bibitem{Felipe2011}  J. C. C. Felipe, L. C. T. Brito, M. Sampaio and M. C. Nemes, Phys.  Lett.  B700 (2011) 86.
	
	\bibitem{Cherchiglia2012}  A. L. Cherchiglia, L. A. Cabral, M. C. Nemes and M. Sampaio,  Phys. Rev. D87 (2013) 065011. 
	
	\bibitem{Felipe2014} J. C. C. Felipe, A. R. Vieira, A. L. Cherchiglia, A. P. Ba\^eta Scarpelli and M. Sampaio, Phys. Rev. D89 (2014) 105034.
	\bibitem{JOILSON} J. Porto, B. Hiller, A. L. Cherchiglia and Marcos Sampaio, Eur. Phys. J. C78 (2018) 160.
	\bibitem{ZURICH} C. Gnendiger et al., Eur. Phys. J. C77 (2017) 471.
	\bibitem{GNEDIGER} C. Gnendiger, A. Signer and D. Stockinger, Phys. Lett. B733 (2014) 296.
	\bibitem{FAZIO} R. A. Fazio, P. Mastrolia, E. Mirabella and W. J. Torres Bobadilla, Eur. Phys. J. C74 (2014) 3197.
	\bibitem{WILLIAM} W. J. T. Bobadilla, PoS RADCOR (2016) 2015.
	\bibitem{PITTAU} R. Pittau, JHEP 1211 (2012) 151.
	\bibitem{ROGER} R. J. Hernandez-Pinto, G. F. R. Sborlini and G. Rodrigo, JHEP 1602 (2016) 044.
	\bibitem{GERMAN} G. F. R. Sborlini, F. Driencourt-Mangin, R. Hernandez-Pinto and G. Rodrigo, JHEP 1608 (2016) 160.
	\bibitem{DWITT} B. De Wit and D. Z. Freedman, Phys. Rev. D12 (1975) 2286.
	\bibitem{ABBOTT1}  L. F. Abbott, M. T. Grisaru and H. J. Schnitzer, Phys. Lett. B71 (1977) 161.
	
  
\bibitem{CORIANO} 
  C.~Coriano, A.~Costantini, L.~Delle Rose and M.~Serino,
  JHEP 1406 (2014) 136.
  
\bibitem{MehtaN}
  M.~R.~Mehta,
  Phys.\ Lett.\ B274 (1992) 53.     
  	
	
	
	
	\bibitem{ABBOTT2} L. F. Abbott, M. T. Grisaru and H. J. Schnitzer, Phys. Rev. D16 (1977) 2995.
	\bibitem{MACKEON} A. M. Chowdhury, V. Elias and D. G. C. McKeon, Phys. Rev. D34 (1986) 619 ;  V. Elias, G. McKeon, S. B. Phillips and R. B. Mann,  Can. J. Phys. 63 (1985) 1453.
	\bibitem{JACKIWFU}  R. Jackiw, Int. J. Mod. Phys. B14 (2000) 2011.
	\bibitem{BPHZ} N. N. Bogoliubov and O. S. Parasiuk, Acta Math. 97 (1957) 227; O. S. Parasiuk, Ukr. Mat. Zh.  12 (1960) 287; K. Hepp, Commun. Math.  Phys. 2 (1966) 301; W. Zimmermann, Commun. Math. Phys.  15 (1969) 208.
	\bibitem{ADRIANOPEREZ} A. M. Bruque, A. Cherchiglia, Manolo Perez-Victoria, arXiV:1803.09764 [hep-ph].	
	\bibitem{YU} H. L.-Yu and W. B. Yeung, Phys. Rev. D35 (1987) 3955.
	\bibitem{TSAI} Er-Cheng Tsai, Phys. Rev. D83 (2011) 065011 ; Phys. Rev. D83 (2011) 025020 ; arXiV : 0905.1479.
	\bibitem{FERRARI} R. Ferrari, arXiv : 1403.4212.
	\bibitem{BONNEAU} G. Bonneau, Phys.Lett. B96 (1980) 147 ; Nuc. Phys. B177 (1981)523.
	\bibitem{ELIAS} V. Elias, G. McKeon and R. B. Mann, Nuc. Phys. B229 (1983) 487.
	\bibitem{MARTIN} C. P. Martin and D. Sanchez-Ruiz, Nuc. Phys. B572 (2000) 387.
	\bibitem{CYNOLTER} G. Cynolter and E. Lendvai, Mod. Phys. Lett. A26 (2011) 1537 ; Central Eur. J. Phys. 9 (2011) 1237.
	\bibitem{DALLABONA} O. A. Battistel, M. V. S. Fonseca and G. Dallabona, Phys. Rev. D85 (2012) 085007 ; O. A. Battistel and G. Dallabona, J. Phys. G28 (2002) 2539.
	\bibitem{FUJIKAWA} K. Fujikawa and H. Suzuki, Oxford, UK : Clarendon (2004).
	\bibitem{ZEE} B. Zumino, Y.-S. Wu and A. Zee, Nuc. Phys. B239 (1984) 477.
	\bibitem{BARDEENOV} W. A. Bardeen in ``{\it{Anomalies}}", Lec. Not. Phys. 558 (2000) 3.
	\bibitem{CURRENT} S.B. Treiman, E. Witten, R. Jackiw and B. Zumino, {\textit{Current Algebra And Anomalies}}, Singapore: World Scientific (1985).	
	\bibitem{JEGERLEHNER} F. Jegerlehner, Eur. Phys. J. C18 (2001) 673.
	\bibitem{SIGNER5} C. Gnendiger and A. Signer, Phys. Rev. D97 (2018) 096006. 

	
	\bibitem{MAJUMDAR} P. Majumdar, E. C. Poggio and H. J. Schnitzer, Phys. Lett. B93 (1980) 321.
	\bibitem{KUMAR} S. Kumar and Y. Fujii, Prog. Theor. Phys. 68 (1982) 294.
	\bibitem{INAGAKI} H. Inagaki, Phys. Lett. B77 (1978) 56.
	\bibitem{HAGIWARA} T. Hagiwara and So-Young Pi, Ann. Of Phys. 130 (1980) 282.
	\bibitem{NICOLAI} H. Nicolai and P. K. Townsend, Phys. Lett. B93 (1980) 111.
	\bibitem{CURTRIGHT} T. Curtright, Phys. Lett. B71 (1977) 185.
	\bibitem{PEREZIRR} M. Perez-Victoria, JHEP 0104 (2001) 032.
	
	\bibitem{SMAILAGIC} A. Smailagic, J. Phys. A : Math. Gen. 17 (1984) 725.
	\bibitem{MEHTA} M. R. Mehta, Phys. Rev. D44 (1991) 3303.
	\bibitem{DALLABONA2} O. A. Battistel and G. Dallabona, Eur. Phys. J. C45 (2006) 721.
	
\end{thebibliography}
\end{document}